\documentclass[sigconf,nonacm,screen]{acmart}
\newcommand{\mode}{final}
\usepackage{ifthen}

\usepackage{bbm}
\usepackage{amsfonts}
\usepackage{amsmath}
\usepackage{mathtools}
\usepackage{bm} %
\usepackage{physics}

\hypersetup{
    colorlinks=true,
    linkcolor=blue,
    filecolor=magenta,
    urlcolor=cyan,
}
\usepackage{cleveref} %
\usepackage{tablefootnote}
\usepackage{autonum}
\crefformat{footnote}{#2\footnotemark[#1]#3}
\crefname{equation}{}{}
\creflabelformat{equation}{#2Equation #1#3}
\crefname{figure}{}{}
\creflabelformat{figure}{#2Figure #1#3}
\crefname{table}{}{}
\creflabelformat{table}{#2Table #1#3}
\crefname{algorithm}{}{}
\creflabelformat{algorithm}{#2Algorithm #1#3}
\crefname{listing}{}{}
\creflabelformat{listing}{#2List #1#3}
\crefname{code}{}{}
\creflabelformat{code}{#2Code #1#3}
\crefname{chapter}{}{}
\creflabelformat{chapter}{#2\S#1#3}
\crefname{section}{}{}
\creflabelformat{section}{#2\S#1#3}
\crefname{subsection}{}{}
\creflabelformat{subsection}{#2\S#1#3}

\newcommand{\footnotetextx}[2]{%
    \stepcounter{footnote}%
    \newcounter{#1}%
    \setcounter{#1}{\value{footnote}}%
    \footnotetext[\value{#1}]{%
        \addtocounter{footnote}{-1}%
        \refstepcounter{footnote}\label{footnote:#1}#2%
    }%
}
\newcommand{\footnotemarkx}[1]{%
    $^{\ref{footnote:#1}}$%
}

\usepackage{bibentry}

\usepackage{url}
\usepackage{diagbox} %
\usepackage{multirow}
\usepackage{tabularx}
\newcommand{\rowsl}[2]{\begin{tabular}{l}#1 \\ #2\end{tabular}}

\usepackage{algorithm}
\usepackage{algpseudocode}
\algnewcommand\algorithmicinput{\textbf{Input:}}
\algnewcommand\INPUT{\item[\algorithmicinput]}
\algnewcommand\algorithmicoutput{\textbf{Output:}}
\algnewcommand\OUTPUT{\item[\algorithmicoutput]}
\algnewcommand\algorithmicforeach{\textbf{for}}
\algdef{S}[FOR]{ForEach}[1]{\algorithmicforeach\ #1\ \algorithmicdo}
\algnewcommand\algorithmicparallelforeach{\textbf{parallel for}}
\algdef{S}[FOR]{PForEach}[1]{\algorithmicparallelforeach\ #1\ \algorithmicdo}

\usepackage{amsthm}
\theoremstyle{definition}

\newtheorem{thm*}{Theorem}
\crefname{thm}{}{}
\crefname{thm*}{}{}
\creflabelformat{thm}{#2Theorem #1#3}
\creflabelformat{thm*}{#2Theorem #1#3}

\newtheorem{dfn*}{Definition}
\crefname{dfn}{}{}
\crefname{dfn*}{}{}
\creflabelformat{dfn}{#2Definition #1#3}
\creflabelformat{dfn*}{#2Definition #1#3}

\newtheorem{cor*}{Corollary}
\crefname{cor}{}{}
\crefname{cor*}{}{}
\creflabelformat{cor}{#2Corollary #1#3}
\creflabelformat{cor*}{#2Corollary #1#3}
\newtheorem{lem}{Lemma}
\newtheorem{lem*}{Lemma}
\crefname{lem}{}{}
\crefname{lem*}{}{}
\creflabelformat{lem}{#2Lemma #1#3}
\creflabelformat{lem*}{#2Lemma #1#3}

\usepackage{soul}

\usepackage{cuted}
\usepackage{multicol}

\usepackage{graphicx} %
\usepackage{color} %
\usepackage{xcolor}
\usepackage{here}
\usepackage{ragged2e}
\usepackage{subcaption}
\usepackage{caption}
\let\oldenumerate\enumerate
\renewcommand{\enumerate}{
    \oldenumerate
    \setlength{\itemsep}{1.2pt}
    \setlength{\itemindent}{0pt}
    \setlength{\parskip}{0pt}
    \setlength{\parsep}{0pt}
}
\let\olditemize\itemize
\renewcommand{\itemize}{
    \olditemize
    \setlength{\itemsep}{1.2pt}
    \setlength{\itemindent}{0pt}
    \setlength{\parskip}{0pt}
    \setlength{\parsep}{0pt}
}

\renewcommand{\paragraph}{}
\newcommand{\xparagraph}[1]{\paragraph{\textbf{#1: }}}

\newcommand{\eqlabel}[1]{\refstepcounter{equation}\quad\text{(\theequation)}\label{#1}}

\usepackage[\mode]{fixme}
\fxsetup{theme=color,inline=true,margin=false}
\definecolor{fxtarget}{rgb}{1,0,0}
\makeatletter
\renewcommand*\FXLayoutInline[3]{{\@fxuseface{inline}\textcolor{fxtarget}{[#2]}}}
\renewcommand*\FXTargetLayoutColor[2]{{\@fxuseface{inline}\textcolor{purple}{[#2]}}\textcolor{fxtarget}{$\leftarrow$}}
\makeatother
\newcommand*{\memo}[1]{\fxwarning{#1}}

\usepackage[cal=boondoxo]{mathalfa}

\newcommand{\bbE}[0]{\mathbb{E}}
\newcommand{\bbOne}[0]{\mathbbm{1}}
\newcommand{\mcC}[0]{\mathcal{C}}
\newcommand{\mcS}[0]{\mathcal{S}}

\newcommand{\mcM}[0]{\mathcal{M}}
\newcommand{\mcE}[0]{\mathcal{E}}
\newcommand{\mce}[0]{\mathcal{e}}

\newcommand{\mcf}[0]{\mathcal{f}}
\newcommand{\mbcf}[0]{\textbf{f}}

\newcommand{\bA}[0]{\bm{A}}
\newcommand{\bT}[0]{\bm{T}}

\newcommand{\bJ}[0]{\bm{J}}
\newcommand{\bF}[0]{\bm{F}}

\newcommand{\FA}[1]{{}^{\forall}#1}
\newcommand{\EX}[1]{{}^{\exists}#1}

\newcommand{\SB}[2]{\scalebox{#1}[1]{#2}}

\newcommand\vldbavailabilityurl{https://github.com/OnizukaLab/Scardina}

\title{Scardina: Scalable Join Cardinality Estimation \\ by Multiple Density Estimators}
\author{Ryuichi Ito}
\affiliation{%
  \institution{Osaka University}
  \city{Osaka}
  \state{Japan}
}
\email{ito.ryuichi@ist.osaka-u.ac.jp}

\author{Yuya Sasaki}
\affiliation{%
  \institution{Osaka University}
  \city{Osaka}
  \state{Japan}
}
\email{sasaki@ist.osaka-u.ac.jp}

\author{Chuan Xiao}
\affiliation{%
  \institution{Osaka University}
  \city{Osaka}
  \state{Japan}
}
\email{chuanx@ist.osaka-u.ac.jp}

\author{Makoto Onizuka}
\affiliation{%
  \institution{Osaka University}
  \city{Osaka}
  \state{Japan}
}
\email{onizuka@ist.osaka-u.ac.jp}

\begin{document} \sloppy 

\ifthenelse{\equal{\mode}{draft}}{
\listoffixmes
\tableofcontents
\listoftables
\listoffigures
}{}

\begin{abstract}
In recent years, machine learning-based cardinality estimation methods are replacing traditional methods.
This change is expected to contribute to one of the most important applications of cardinality estimation, the query optimizer, to speed up query processing.
However, none of the existing methods do not precisely estimate cardinalities when relational schemas consist of many tables with strong correlations between tables/attributes.
This paper describes that multiple density estimators can be combined to effectively target the cardinality estimation of data with large and complex schemas having strong correlations.
We propose Scardina, a new join cardinality estimation method using multiple partitioned models based on the schema structure.
\end{abstract}

\maketitle

\ifdefempty{\vldbavailabilityurl}{}{
\vspace{.3cm}
\begingroup\small\noindent\raggedright\textbf{Artifact Availability:}\\
The source code, data, and/or other artifacts have been made available at \url{\vldbavailabilityurl}.
\endgroup
}

\section{Introduction \label{sec:intro}}
Cardinality estimation is a fundamental task in database systems that estimates the number of records in a database that satisfy a given query condition.
Especially, join cardinality estimation is used in query optimizers and is known to significantly impact query processing performance~\cite{cardest}.
The scale of data and schemas to be queried and processed is snowballing, and it is necessary to support such large-scale databases~\cite{largejoin,largejoinbi}.

Traditionally, cardinality estimation uses statistical information such as histograms and is widely implemented in real-world database systems~\cite{pgcardest,mariadb}.
These traditional approaches utilize a distribution of data independent of each attribute.
However, data is generally biased and correlated across attributes, resulting in a lossy independence assumption, which leads to low estimation performance.
The low estimation performance prevents query optimization from achieving high query processing performance.
To solve this problem, machine learning methods have been proposed in recent years.
There are two main approaches to estimating join cardinality using machine learning.
One is a query-driven approach~\cite{learnedcardinalities, lightweight} that treats cardinality estimation as a regression problem with queries as input and cardinalities as output.
However, the query-driven approach is challenging when the data is significant because it is difficult to collect a large number of training labels.
The other is a data-driven approach~\cite{neurocard, deepdb, flat} that learns data distribution from the database and estimates cardinalities by density estimation for given query conditions.
The data-driven approach achieves high estimation accuracy by precisely capturing correlations among attributes.
However, when the schema is large, i.e., the larger number of tables, it becomes difficult to construct models and estimation accuracy degrades due to the estimation of a large number of joins~\cite{kyoungmin}.

\memo{Add details of our method}
In this paper, we propose \setul{1pt}{0.4pt}\ul{\textbf{S}}\setul{1pt}{0.4pt}\underline{\ul{\textbf{ca}}}\setul{3pt}{0.4pt}\ul{\textbf{rdina}}, a \setul{1pt}{0.4pt}\ul{\textbf{sca}}lable join \setul{3pt}{0.4pt}\ul{\textbf{cardina}}lity estimation method for large schemas.
We focus on the fact that most queries access a small subset of  tables in the schema. This phenomenon is generally true in particular when the schema contains a large number of tables. 
By leveraging the fact, existing methods, such as DeepDB, partition a schema into small subschemas and build a single model for each subschema so that the query optimizer can estimate cardinalities by filtering out the subschemas that a given query does not refer to; DeepDB utilizes the models built only from the tables a given query refers to.
However, DeepDB assumes no correlation between different subschemas, which causes poor performance for queries crossing multiple subschemas. And DeepDB is not practical for cases with a large number of tables and attributes because of the high cost of partitioning schemas.
In order to mitigate this problem, Scardina employs correlations between subschemas by propagating biased distributions of one model to another model using a biased sampling technique.

The main contributions of this paper are as follows:
\begin{itemize}
    \item We propose Scardina, an efficient method for estimating join cardinalities using multiple density estimators. Scardina has the following three features:
    \begin{itemize}
        \item \textbf{Scalability with respect to schema size}: Lightweight structure-based schema partitioning is performed, and a density estimator is trained for each subschema. Since each density estimator has largely fewer target tables than the total, its model size is kept small. Additionally, since there are no dependencies between density estimators, they can be trained in parallel.
        \item \textbf{Scalability with respect to data size}: By appropriately sampling the data used for training the density estimator, Scardina can flexibly handle small to large-scale data.
        \item \textbf{High estimation accuracy}: Because Scardina captures and maintains the distribution of data for each partitioned subschema, it can respond to queries with a minimum set of estimators, resulting in high estimation accuracy. This property is particularly effective when used with the query optimizer. The estimation accuracy is maintained even when inference is performed across subschemas by sharing intermediate results between partitions.
    \end{itemize}
    \item We evaluate Scardina using real-world benchmarks and confirm that Scardina can efficiently estimate the join cardinality and contribute to faster query processing via the query optimizer. %
\end{itemize}

The structure of this paper is as follows:
In \cref{sec:preliminaries}, we introduce the necessary definitions for this paper, and in \cref{sec:relworks}, we introduce related research and discuss its issues.
In \cref{sec:prop}, we propose a new join cardinality estimation method. %
We evaluate the methods and identify the strengths and weaknesses of each method in \cref{sec:eval}.
Finally, we conclude this paper in \cref{sec:conc}.

\section{Preliminaries \label{sec:preliminaries}}
This section introduces the definitions and problem formulation necessary to deal with join cardinality estimation in a data-driven manner.

\subsection{Notation \label{ssec:notation}}
The definitions of the main symbols are given in \cref{tbl:notation}, and the database schema, density estimator and how to transform the query into a graph structure are also described.

\begin{table*}[ht]
\caption{Notation \label{tbl:notation}}
\begin{tabularx}{\linewidth}{l|X}
    Symbol & Definition \\ \hline \hline
    $A$ &
        Attribute. \\ \hline
    $V$ &
        $v \in V$. Set of vertices representing tables. $v$ denotes a table in schema graphs, also denoted as $T \in \bT$. \\ \hline
    $E$ &
        $(u, v, c) \in E, c:=(c_u, c_v)$.
        Set of labeled directed edges connecting vertices $u$ and $v$.
        It represents foreign key constraint that table $u$ and $v$ are in one-to-many relationship with $u.c_u = v.c_v$. \\ \hline
    $G$ &
        $G := (V,E)$.
        Directed acyclic multigraph with edge labels constructed from set of vertices $V$ and set of edges $E$.
        It represents schema of a database. For simplicity, we assume there are no self-loops. \\ \hline
    $G_{GLB}$ &
        $G_{GLB} := (V_{GLB},E_{GLB})$.
        Directed acyclic multigraph with edge labels that represents schema of entire database. \\ \hline
    $G_Q$ &
        Directed acyclic graph that represents tables and relationships targeted by query $Q$.
        When viewed as undirected graph, connected.
        Also, subgraph of schema graph of entire database, i.e., $G_Q \subseteq G_{GLB}$.
        In this paper, unless otherwise noted, directed tree. \\ \hline
    $R_Q$ &
        Predicate conditions.
        $R_Q(A)$ represents range of domain of attribute $A$ that satisfies predicate condition of query $Q$, i.e., $R_Q(A) \subseteq \text{dom}(A)$.
        If no predicate is specified, it is equal to original domain, i.e., $R_Q(A) = \text{dom}(A)$.
        Range that satisfies all predicate conditions of $Q$ can be denoted as $\prod_{A}{R_Q(A)}$, which is hyper-rectangle.
        If $Q$ is clear from context, it is denoted as $R$. \\ \hline
    $Q$ &
        $Q:=(G_Q,R_Q)$.
        Query composed of query graph $G_Q$ and predicate conditions $R_Q(A)$. \\ \hline
    $\text{dom}(A)$ &
        Domain of attribute $A$. \\ \hline
    $J_V$ &
        Table with fully outer join of vertex set $V$. \\ \hline
    $N_T$ &
        Table flag.
        Special attribute indicates inclusion of table $T$. Table flag will be explained in \cref{ssec:relwork_univrel}. \\ \hline
    $F$ &
        Fanout.
        Special attribute indicates how many tuples it corresponds to. Fanout will be explained in \cref{ssec:relwork_univrel}. \\ \hline
    $\mcM_T$ &
        Density estimator for table $T$. \\ \hline
    \rowsl{$P_{T}(a_1 \in R(A_1),$}{$\dots, a_n \in R(A_n))$} &
        Joint probability that attributes $A_1,\dots,A_n$ satisfy conditions in table $T$.
        If target is clear, denoted as $P(a_1 \in R(A_1),\dots,a_n \in R(A_n))$.
        If specific conditions are omitted, assumed to indicate joint distribution: $P_{T}(A_1,\dots,A_n)$. \\ \hline
    $\mcC(Q)$ &
        Cardinality based on query $Q$. \\ \hline
    $\mcS(Q)$ &
        Selectivity based on query $Q$. \\ \hline
    $\hat{\cdot}$ &
        Estimated value obtained from machine learning models or statistical information. \\ \hline
    $\bbOne_{c}$ &
        Indicator function that takes 1 if condition $c$ is satisfied, and 0 otherwise.\\ \hline
    $G[S]$ &
        Subgraph induced by set of vertices $S$ from graph $G$. \\ \hline
    \rowsl{$N^{+}_{G}[v], N^{-}_{G}[v],$}{$N^{-}_{G}(v), N^{+}_{G}(v)$} &
        Neighbors of vertex $v$ in Graph $G$.
        $N^+$ and $N^-$ indicate out-neighborhood and in-neighborhood, respectively.
        $[v]$ and $(v)$ indicate closed neighborhood and open neighborhood, respectively. \\ \hline
    $H$ &
        $H:=(V,\mcE)$.
        Multihypergraph consisting of vertex set $V$ and hyperedge set $\mcE$.
        In this paper, it represents subschemas on global schema graph. \\ \hline
    $\underset{t \sim \mcM}{\bbE}\lbrack \cdot \rbrack$ &
        Expectation based on sample $t$ extracted from distribution $\mcM$. \\ \hline
\end{tabularx}
\end{table*}

\begin{figure}[ht]
    \centering
    \begin{tabular}{cc}
        \begin{minipage}[t]{0.42\columnwidth}
            \centering
            \includegraphics[width=\hsize]{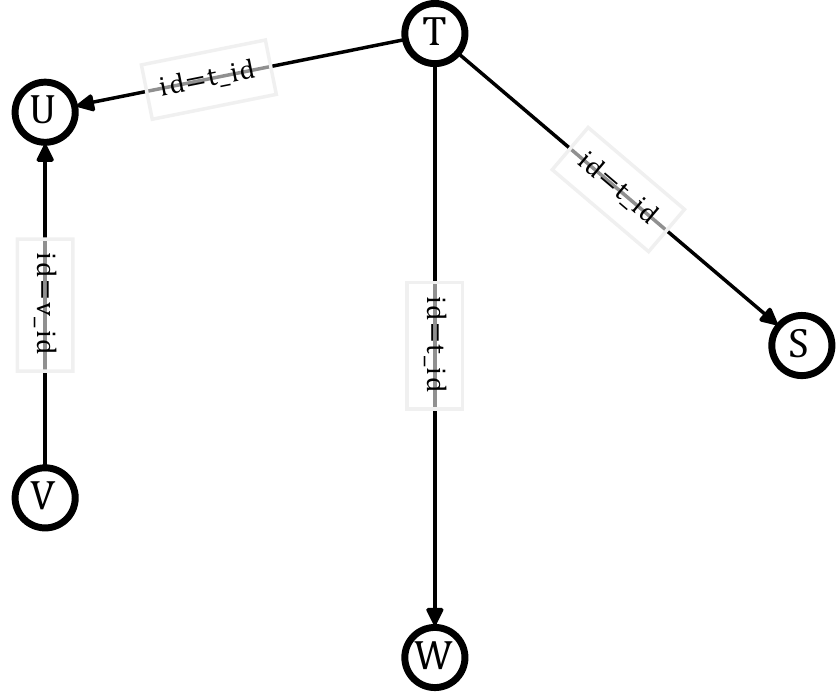}
            \subcaption{Schema graph $G_{GLB}$. Each vertex represents table, each edge represents relationship i.e., foreign key constraint, and each edge label represents content of join constraint. \label{fig:schema_g}}
        \end{minipage}
        \hspace{0.06\columnwidth}
        \begin{minipage}[t]{0.42\columnwidth}
            \centering
            \includegraphics[width=\hsize]{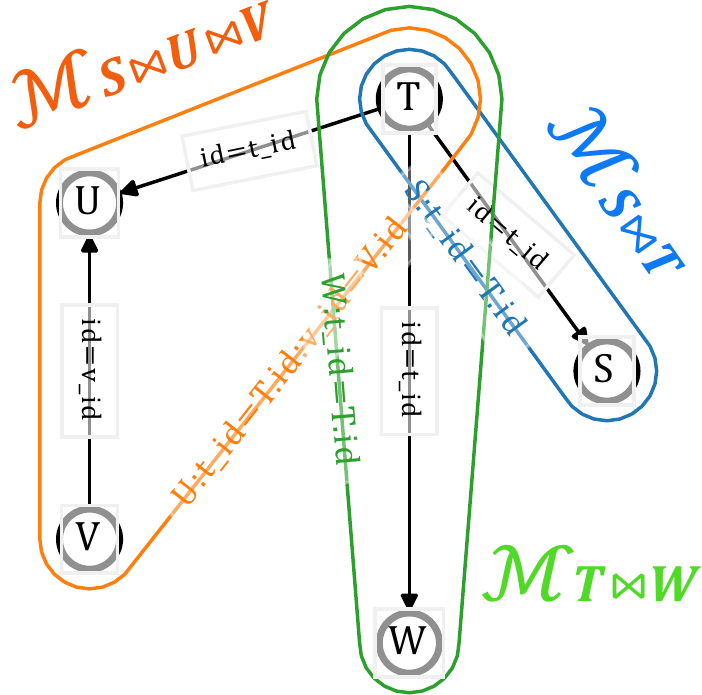}
            \subcaption{Example of correspondence between schema graph $G_{GLB}$ and density estimator $\mcM$. Density estimators are constructed for each hyperedge. Subscript of $\mcM$ represents target table\footnote{Here, $\Join$ denotes a full outer join.}. Shape of hypergraph may differ depending on method used. \label{fig:estimator_g}}
        \end{minipage}
        \\
        \begin{minipage}[t]{0.42\columnwidth}
            \centering
            \includegraphics[width=\hsize]{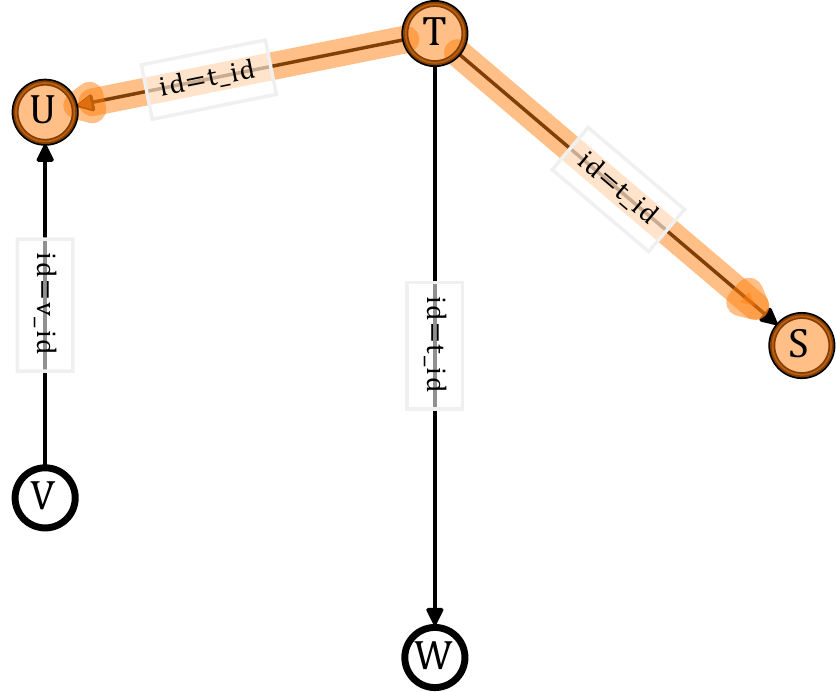}
            \subcaption{Query graph $G_Q$ on schema graph $G_{GLB}$. Orange graph represents query graph. \label{fig:query_g}}
        \end{minipage}
        \hspace{0.06\columnwidth}
        \begin{minipage}[t]{0.42\columnwidth}
            \centering
            \includegraphics[width=\hsize]{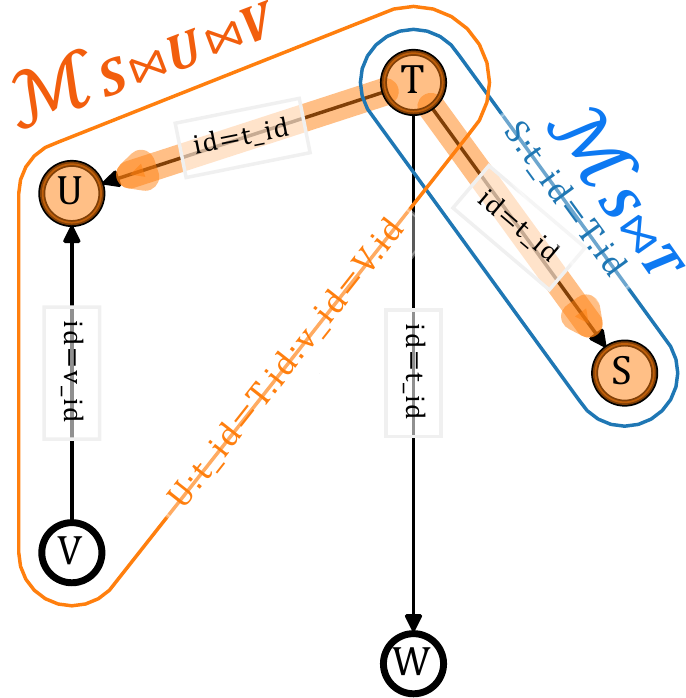}
            \subcaption{Example of correspondence between query graph $G_Q$ and density estimators $\mcM$. Ranges of density estimator vary based on method used. Irrelevant density estimators with respect to query $\mcM_{T \Join W}$ are not selected. \label{fig:estimator_query_g}}
        \end{minipage}
     \end{tabular}
     \caption{Example of graph structures representing schemas and queries. \label{fig:graphs}}
\end{figure}

\xparagraph{Graphs associated with join cardinality estimation}
\cref{fig:graphs} shows examples of the global database schema graph and the query graph used in this paper.
The schema graph $G_{GLB}$, shown in \cref{fig:schema_g}, is represented by $G_{GLB} := (V_{GLB}, E_{GLB})$ using $V_{GLB}=\{S,T,U,V,W\}$ and $E_{GLB}=\{(T,S,(\text{id},\text{t\_id})),\allowbreak (T,U,(\text{id},\text{t\_id})),\allowbreak (T,W,(\text{id},\text{t\_id})),\allowbreak (V,U,(\text{id},\text{v\_id}))\}$.

\cref{fig:estimator_g,fig:estimator_query_g} demonstrate the correspondence of the density estimators to the global schema graph and their usage during inference.
Another perspective is that the correspondence between the schema graphs and the density estimators can be viewed as a hypergraph $H:=(V_{GLB},\mcE)$ with hyperedges $\mcE=\{(S,T),(T,U,V),(T,W)\}$ (\cref{fig:estimator_g}), where each density estimator covers a set of vertices.
It should be noted that the specific range of each density estimator's coverage and its usage during inference may differ depending on the methods.

For example, the query graph $G_Q$ (\cref{fig:query_g}) can be expressed as $G_Q:=(V_Q,E_Q)$ using $V_Q=\{S,T,U\}$ and $E_Q=\{(t,s,(\text{id},\text{t\_id})),\allowbreak (t,u,(\text{id},\text{t\_id}))\}$.

\subsection{Problem Formulation \label{ssec:form}}
In this section, we formulate the join cardinality using data distribution.
We first examine the cardinality of a single table and then address the join cardinality of multiple tables.

\xparagraph{Cardinality of a Single Table}
In the case of a single table, the cardinality $\mcC(Q)$ of a query $Q$ based on the table $T$ can be expressed as the product of the probability that all attributes $\bA$ of the table satisfy the conditions of the query, known as the selectivity $\mcS(Q)$, and the size of the table $|T|$ (\cref{eq:ce_single_c}).
The joint probability can also be expressed as the product of conditional probabilities using the chain rule.
Since many density estimators deal with conditional probabilities of attributes instead of joint probabilities, we often use the estimation of conditional probabilities for cardinality estimation.
\begin{equation}
\label{eq:ce_single_c}
\begin{split}
\mathcal{C}(Q)
    &= |T| \cdot P_{T}(A_1 \in R_{Q}(A_1), \dots, A_n \in R_{Q}(A_n))
\\
    &=
    \begin{aligned}
        |T| &\cdot P_{T}(A_1 \in R_{Q}(A_1)) \\
        &\begin{aligned}\dotsm P_T(&A_n \in R_{Q}(A_n) \mid \\&A_{n-1} \in R_{Q}(A_{n-1}), \dots, A_1 \in R_{Q}(A_1))\end{aligned}
    \end{aligned}
\end{split}
\end{equation}

\xparagraph{Join Cardinality of Multiple Tables}
Let $J_{\bT}$ be the table obtained by performing a full outer join on all tables.
Also, suppose $G_Q = G_{GLB}$, meaning that the query $Q$ accesses all tables (The case $G_Q \subset G_{GLB}$ is discussed in \cref{ssec:relwork_univrel}).
In this case, the join cardinality $\mcC(Q)$ based on the query $Q$ can be calculated as the product of the joint probability satisfying all the attributes $\bA$ in the table $J_{\bT}$ and the number of tuples $|J_{\bT}|$ (\cref{eq:ce_multi_c}).
It can also be expressed as the product of conditional probabilities by the chain rule, as in the single-table case.
\begin{equation}
\label{eq:ce_multi_c}
\begin{split} 
\mathcal{C}(Q)
    &= |J_{\bT}| \cdot P_{J_{\bT}}(A_1 \in R_{Q}(A_1), \dots, A_n \in R_{Q}(A_n))
    \\
    &=
    \begin{aligned}
        |J_{\bT}| &\cdot P_{J_{\bT}}(A_1 \in R_{Q}(A_1), \dots, A_n \in R_{Q}(A_n)) \\
            &\begin{aligned}\dotsm P_{J_{\bT}}(&A_n \in R_{Q}(A_n) \mid \\&A_{n-1} \in R_{Q}(A_{n-1}), \dots, A_1 \in R_{Q}(A_1))\end{aligned}
    \end{aligned}
\end{split}
\end{equation}

Although Scardina can handle arbitrary relationships defined in advance, for the sake of simplicity, this paper only describes joins based on one-to-many relationships with foreign key constraints.
In addition, although inner joins and outer joins are different in join cardinality estimation, this paper only describes inner joins\footnote{Scardina can handle both inner and outer joins in the same way as DeepDB and NeuroCard.}.

\section{Challenges in Large-Scale Schemas \label{sec:relworks}}
\memo{better to make subschema for each method clear (all methods )}

This section explains existing join cardinality estimation methods and the challenges in applying them to data with large-scale schemas.
Existing data-driven join cardinality estimation methods are roughly divided into methods that directly use statistical information and data, such as histograms and sampling, and machine learning methods.
In the following, we subdivide each approach and summarize the outline and problems of each method.

\subsection{Case 1 (Data-driven): Density Estimation by Independent Distributions per Attribute \label{ssec:relwork_hist}}
\begin{figure}[ht]
    \centering
    \begin{minipage}[t]{0.47\columnwidth}
        \centering
        \includegraphics[width=\hsize]{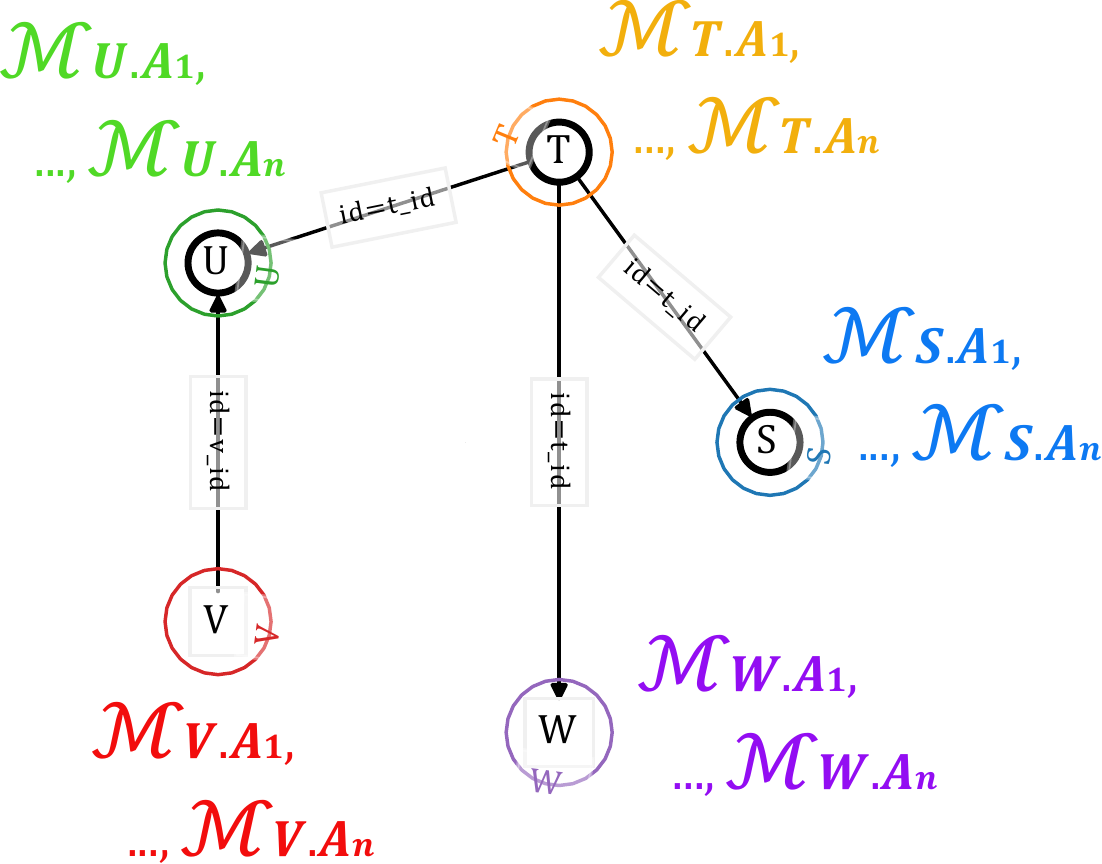}
        \subcaption{Example of correspondence between schema graph and density estimator for statistical information-based methods. Density estimator here is set of histograms. Note that it is constructed per attribute, not per table. \label{fig:g_hist_s}}
    \end{minipage}
    \hspace{0.04\columnwidth}
    \begin{minipage}[t]{0.47\columnwidth}
        \centering
        \includegraphics[width=\hsize]{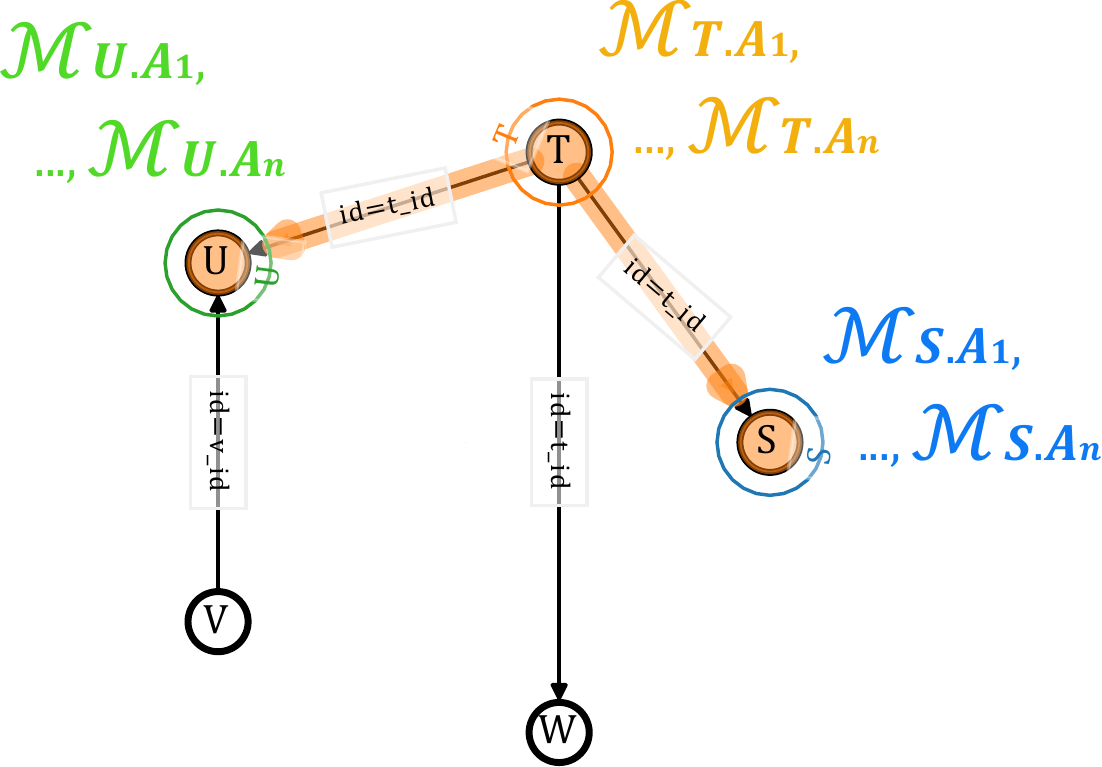}
        \subcaption{Example of correspondence between query graph and density estimator. Among tables in query graph, only histograms for attributes specified in query are used. \label{fig:g_hist_q}}
    \end{minipage}
    \caption{Method based on independent statical information for each attribute \label{fig:g_hist}}
\end{figure}

In many existing database systems, such as PostgreSQL and MariaDB, statistical information is used for the join cardinality estimation~\cite{pgcardest,mariadb}.
As a form of statistical information, histograms are created for each attribute\footnote{In practice, statistical information such as frequent value lists and empirical magic numbers are sometimes used in combination.} (\cref{fig:g_hist}).
In practice, histograms are constructed for each attribute since the number of attribute combinations is vast and product domains must also be handled.
This method has a simple structure and scales even with large data sizes.
During inference, the histograms are used for the density estimation by calculating the percentage of bins that match the query condition.
Since the histograms are independent for each attribute, if the query condition crosses multiple attributes, the correlation between attributes is ignored in the estimation.
In the case of the join cardinality estimation, since there is no cross-table information, the assumption that the joins are uniformly distributed across the key attributes is used.
This strategy can be expressed as \cref{eq:hist_multi_p}\footnote{For simplicity, assuming that the bin width is small enough that the assumption of uniform distribution within the bin has no effect}.
The notation using the expected value is \cref{eq:hist}.
These unrealistic assumptions result in poor estimation performance.
In particular, performance tends to deteriorate when using many predicates or joins.

Scardina uses denoising autoencoders internally to achieve high estimation accuracy without approximations based on the assumption of independence among attributes or uniform distribution as in histogram-based methods.

\begin{align}
\mathcal{C}_{INDEP}(Q)
&\begin{aligned}=
    \prod_{T \in V_Q}{|T|}
    \cdot
    &\prod_{A_i \in \bA}{
        \underset{a \sim \mcM_{A_i}}{\bbE}
        \left\lbrack{
            \bbOne_{a \in R_{Q}(A_i)}
        }\right\rbrack
    }
    \cdot
    \\&\prod_{(T_1, T_2, (A_1, A_2)) \in E_Q}{
        \frac{1}{\text{dom}(T_1.A_1)}
    }
    \label{eq:hist}\end{aligned}
    \\
\mathcal{C}_{INDEP}(Q)
&\begin{aligned}=
    \prod_{T \in V_Q}{|T|}
    \cdot
    &\prod_{A_i \in \bA}{
        P(A_i \in R_{Q}(A_i))
    }
    \cdot
    \\&\prod_{(T_1, T_2, (A_1, A_2)) \in E_Q}{
        \frac{1}{\text{dom}(T_1.A_1)}
    }
    \label{eq:hist_multi_p}\end{aligned}
\end{align}

\subsection{Case 2 (Data-driven): Density Estimation by Distribution of Universal Relation \label{ssec:relwork_univrel}}
In NeuroCard~\cite{neurocard} and AutoEncoderUR~\cite{nar_cardest}, join cardinality estimation is performed using a single large density estimator corresponding to the global schema.
These methods assume a full outer join of all tables in the global schema, namely a universal relation, according to predefined relationships.
These methods use two additional virtual attributes to handle queries where the target set of tables is a proper subset of the set of tables in the global schema at the time of inference.
One is \textit{TableFlag}, a boolean value indicating whether the join tuple contains the corresponding table or not, and the other is \textit{Fanout}, a scale factor for each relationship.
The models learn to estimate the density of universal relationships by including these virtual attributes.
For example, NeuroCard uses an autoregressive model and AutoEncoderUR uses a denoising autoencoder as a density estimator.
During inference, they infer the conditional probability of each attribute based on the given query conditions.
These systems also infer TableFlags and Fanouts.
One infers an additional condition that TableFlags are true to reflect the constraint that the tables in the query graph are present in the tuple.
The other infers Fanouts for tables not in the query graph and uses them for downscaling to counteract the effect of unused tables.

Put a query condition and a condition that tables are included are denoted as \cref{eq:def_prob}, the above inference can be expressed as \cref{eq:neurocard}.
In addition to the product of the overall universal relation size $|J_{\bT}|$ and the joint probability $\underset{t \sim \mcM_{J_{\bT}}}{\mathbb{E}}\left\lbrack{\prod_{A \in \bA}{\bbOne_{t[A] \in R_{Q}(A)}}}\right\rbrack$, the condition of the existence of tables included in the query graph $\prod_{T \in V_Q}{\bbOne_{t[N_T]}}$ and the downscaling by Fanout corresponding to joins $join$ not included in the query graph are taken into account simultaneously.
\cref{eq:neurocard} is a strict expression without approximations, such as column independence assumptions.

In this approach, when the data size is large, instead of all full outer joined tuples, we can draw join samples that are i.i.d. and use them as training data. In practice, Exact Weight Algorithm~\cite{joinsampling} is often used to draw samples.
In the case of a large schema size, the range handled by a single density estimator becomes huge in proportion to the schema size, resulting in poor estimation performance due to the complexity of the data distribution to be captured and requiring considerable computational resources.
In addition, it is known that the larger the difference between the schema graph and the query graph, the worse the performance will be due to the subtractive property of the method reducing the influence of unnecessary tables from the universal relation~\cite{kyoungmin}.
This tendency is especially true when using join cardinality estimation in query optimizers since estimation is performed for each subgraph of the query graph for join ordering.
On the other hand, Scardina uses multiple small density estimators that can be trained in parallel to reduce the training cost and achieve high accuracy in real-world applications such as query optimizers.

\begin{figure}[ht]
    \centering
    \begin{minipage}[t]{0.42\columnwidth}
        \centering
        \includegraphics[width=0.95\hsize]{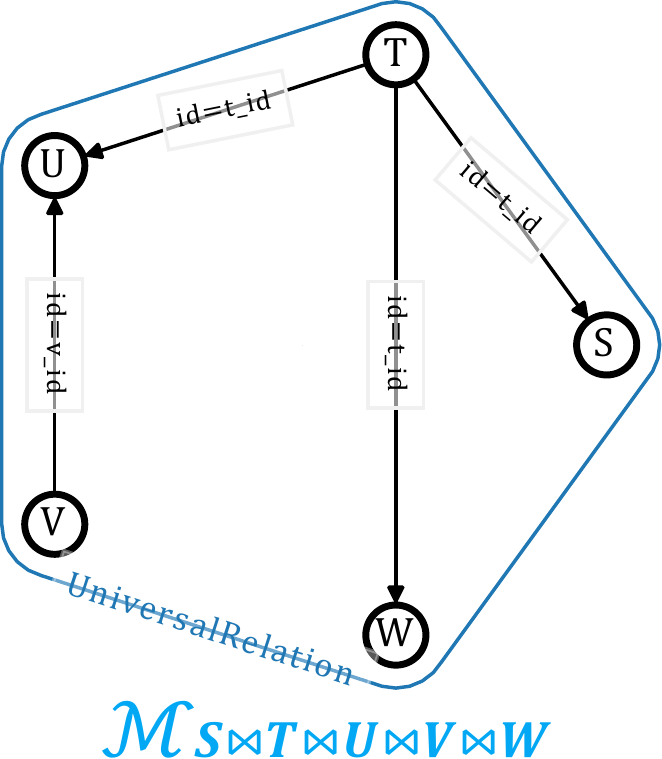}
        \subcaption{Example of correspondence between schema graph and density estimator for universal relation. Universal relation is full outer join of all tables in global schema. \label{fig:g_neurocard_s}}
    \end{minipage}
    \hspace{0.06\columnwidth}
    \begin{minipage}[t]{0.42\columnwidth}
        \centering
        \includegraphics[width=0.95\hsize]{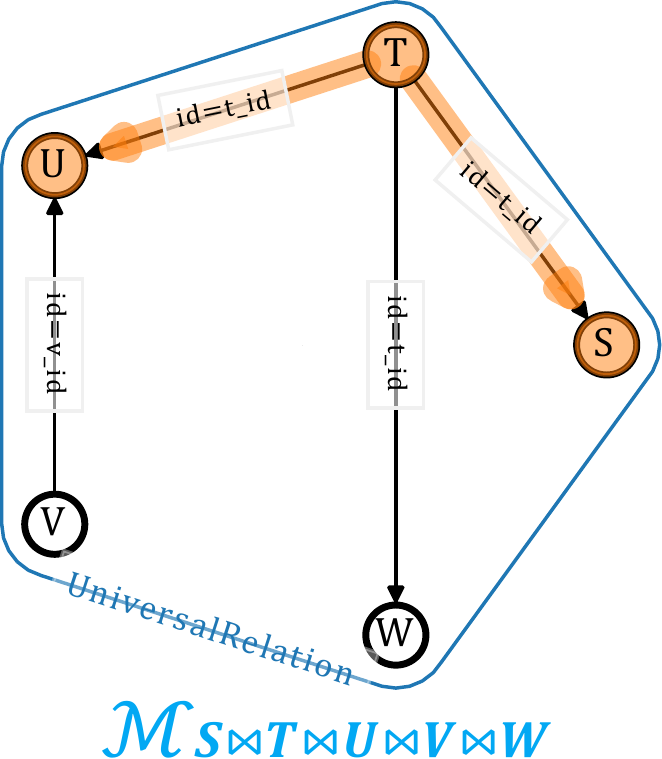}
        \subcaption{Example of correspondence between query graph and density estimator. Single density estimator is always used, requiring downscaling of tables not in query graph. \label{fig:g_neurocard_q}}
    \end{minipage}
    \caption{Method based on universal relation \label{fig:g_neurocard}}
\end{figure}

\begin{align}
\mathbb{I}(\bA, Q, t)
&:=
  \prod_{A \in \bA}{\bbOne_{t[A] \in R_{Q}(A)}}
  \cdot
  \prod_{T \in V_Q}{\bbOne_{t[N_T]}}
\label{eq:def_prob}
\\
\mathcal{C}_{UR}(Q)
&=
    |J_{\bT}|
    \cdot
    \underset{t \sim \mcM_{J_{\bT}}}{\mathbb{E}}
    \left\lbrack{
        \frac{
          \mathbb{I}(\bA, Q, t)
        }{
          \prod_{join \notin E_Q}{t[F_{join}]}
        }
    }\right\rbrack \label{eq:neurocard}
\end{align}

\subsection{Case 3 (Data-driven): Density Estimation by Distributions of Correlation-based Partitioned Subschemas \label{ssec:relwork_corrpart}}
In DeepDB~\cite{deepdb}, the join cardinality estimation is performed by constructing density estimators for each schema partitioned based on correlations between tables.
First, this approach calculates the correlations between tables based on the relationships defined in the schema graph.
DeepDB uses Randomized Dependence Coefficients (RDCs)~\cite{rdc} as correlations.
The calculated RDCs are compared with a manually set threshold; if the values exceed the threshold, the tables are considered correlated; otherwise, they are treated as independent.
The method assumes and handles full outer joined tables, correlation-based joined tables, for each set of tables that are determined to be correlated.
Note that if the threshold is set extremely low, all tables will eventually be combined into a universal relation as described in \cref{ssec:relwork_univrel}; conversely, if set extremely high, all tables become independent.
Therefore, to exploit the advantages of correlation-based partitioning, it is necessary to set a threshold that moderately partitions the schema.
Then, as in \cref{ssec:relwork_univrel}, density estimators are trained to correspond to the correlation-based joined tables, including TableFlags and Fanouts.
DeepDB uses Sum-Product Networks~\cite{sumproductnetworks} as density estimators.

During inference, the join cardinality estimation is performed by traversing the density estimators based on the given query conditions.
When traversing across the density estimators, the increase in cardinality due to the join with the subsequent tables is emulated by upscaling with Fanouts estimation, based on the conditions for the preceding tables.
Estimations of the subsequent density estimators are treated as independent.

In this approach, the schema partitioning is such that there is no overlap between subschemas, and all tables are covered: $({}^{\forall}(J_1, J_2) \in \bm{J}^{2} \land (V_{J_1} \cap V_{J_2}) = \emptyset) \land (\bigcup_{J \in \bJ}{V_J} = V_{GLB})$.
Let the set of correlation-based joined tables required for query $Q$ be $\bJ_Q = \{J \in \bJ \mid V_J \cap V_Q \ne \emptyset\}$, and let $J_1$ be any table initially extracted from $\bJ_Q$.
The join cardinality estimation can be expressed as \cref{eq:deepdb} using \cref{eq:def_prob}.
The cardinality is initially $|J_1|$ and is the product of the expected values of the following four terms for each correlation-based partial join table $J$:
(1) the condition of the query predicate $\prod_{A \in \bA}{\bbOne_{t[A] \in R_{Q}(A)}}$,
(2) the condition that tables are included in the query graph $\prod_{T \in V_Q}{\bbOne_{t[N_T]}}$,
(3) upscaling by joining subsequent tables $\prod_{\substack{(T_1, T_2, (A_1, A_2)) \in E_Q \\ \land T_1 \in V_{J_i} \land T_2 \notin V_{J_i} \land T_2 \in V_{J_{>i}}}}{t[F_{(T_1, T_2, (A_1, A_2))}]}$, and
(4) downscaling to counteract the effect of unused tables $\prod_{join \notin E_Q \land join \in E_{J_i}}{t_i[F_{join}]}$.
The fact that the expected values $\bbE_{t \sim \mcM_J}$ are separated for each correlation-based joined table $J$ means that the estimation includes an approximation that assumes independence.

Compared to the subtractive nature of the universal relation-based approach, this approach exhibits the additive nature\footnote{When viewed within a correlation-based joined table, it has the subtractive nature.}.
Small query graphs are estimated by a small number of density estimators and are not affected by partitioning to any great extent.
Conversely, as the query graph grows, more density estimators are needed for inference, and it is known that upscaling and estimation assuming independence increase, which can easily degrade performance~\cite{kyoungmin}.
One challenge is that RDC has a very large time complexity $o(rn)$ and spatial complexity $o(rn^2)$, assuming an $n$-row $r$-column table.
Simplification is also difficult since coarse correlation decisions can lead to assumptions that do not fit the data and cause estimation errors.

Scardina avoids huge universal relations by partitioning the schema in a manner similar to DeepDB.
Scardina differs from DeepDB in that it performs schema partitioning by referring only to the structure of a schema graph, not data, and does not use manual thresholds.
This makes schema partitioning very inexpensive, does not require manual adjustment, and is scalable with respect to schema size.
In addition, subschemas have overlapped tables so that inference results can be shared among density estimators during inference.
Compared to DeepDB, which treats density estimators independently, Scardina improves performance by using correlation across subschemas.

\begin{figure}[ht]
    \centering
    \begin{minipage}[t]{0.42\columnwidth}
    \centering
        \includegraphics[width=\hsize]{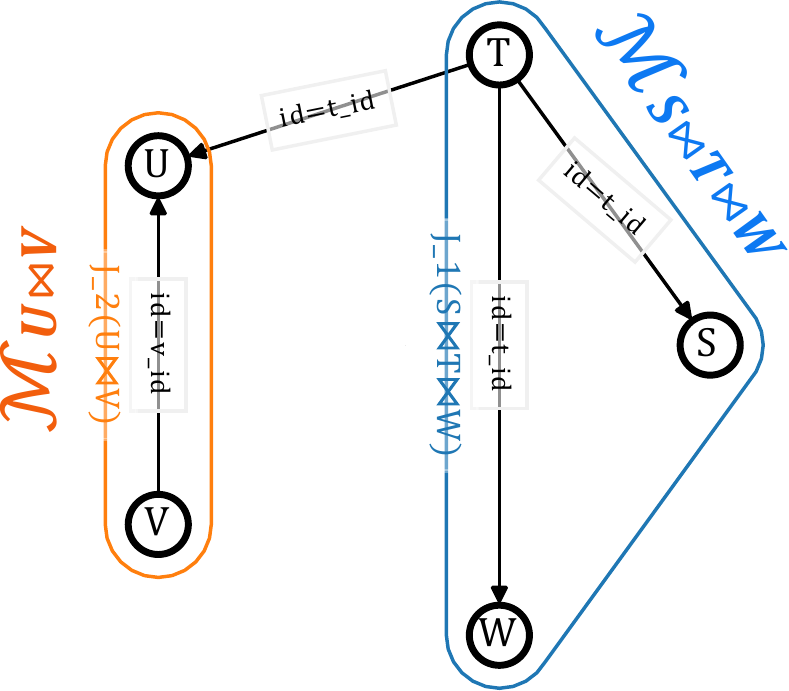}
        \subcaption{Example of correspondence between schema graph and density estimators for correlation-based subschemas. Assuming that $T$ and $U$ are determined to be uncorrelated in precomputation. \label{fig:g_deepdb_s}}
    \end{minipage}
    \hspace{0.06\columnwidth}
    \begin{minipage}[t]{0.42\columnwidth}
    \centering
        \includegraphics[width=\hsize]{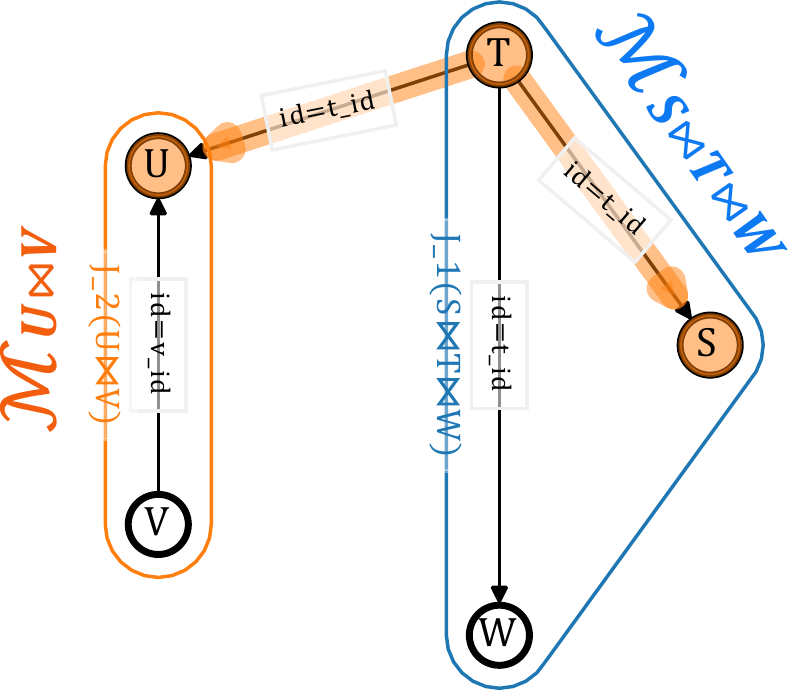}
        \subcaption{Example of correspondence between query graph and density estimators. Density estimators of correlation-based subschemas included in query graph and that are not redundant are used. \label{fig:g_deepdb_q}}
    \end{minipage}
    \caption{Method based on subschemas partitioned based on correlation \label{fig:g_deepdb}}
\end{figure}

\begin{table*}[bt]\begin{tabular}{c}
$\begin{aligned}
\mathcal{C}_{CBP}(Q)
&=
  |J_1|
  \cdot
  \prod_{i=1}^{N}{
    \underset{t \sim \mcM_{J_i}}{\bbE}
    \left\lbrack{
      \frac{\textstyle
        \mathbb{I}(\bA, Q, t)
        \cdot
        \prod_{\substack{(T_1, T_2, (A_1, A_2)) \in E_Q \\
    \land T_1 \in V_{J_i} \land T_2 \notin V_{J_i} \land T_2 \in V_{J_{>i}}}}{
            t[F_{(T_1, T_2, (A_1, A_2))}]}
      }{
      \textstyle\raisebox{-3pt}{$
        \prod_{join \notin E_Q \land join \in E_{J_i}}{t_i[F_{join}]}
      $}
      }
    }\right\rbrack
    }
  \eqlabel{eq:deepdb}
\end{aligned}$
\end{tabular}\end{table*}

\subsection{Case 4 (Workload-driven): Regression \label{ssec:relwork_reg}}
In methods represented by MSCN~\cite{learnedcardinalities,deepsketches}, join cardinality estimation is treated as a regression.
Queries are taken as input and cardinalities are output through models.
By carefully embedding input queries, the join cardinality can also be estimated.
However, there exists a dilemma that we must obtain cardinalities by actually processing queries as labels for training.
As a result, there is a practical challenge of difficulty in increasing training data to improve the estimation performance of models.

\section{Scardina: Scalable Join Cardinality Estimator \label{sec:prop}}
We propose Scardina, a cardinality estimation method using density estimators for each partition based on the structure of the schema graph.
First, we introduce an overview of Scardina and then describe the details of Scardina in \cref{ssec:prop_partition} and later.

\begin{figure}[ht]
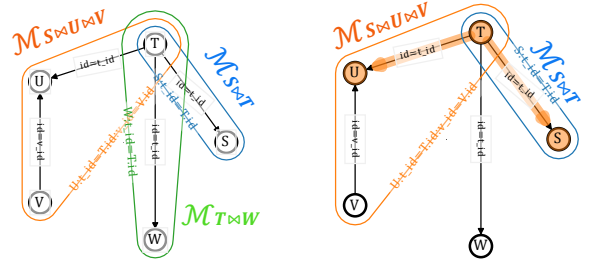

\small
\centering
\begin{minipage}[t]{0.495\columnwidth}
    \centering
    \includegraphics[width=0.8\hsize]{figures/draft/ours_base.pdf}
    \subcaption{Example of correspondence between schema graph and density estimators for each closed in-neighborhood subschema. \label{fig:g_prop_s}}
\end{minipage}
\begin{minipage}[t]{0.495\columnwidth}
    \centering
    \includegraphics[width=0.8\hsize]{figures/draft/ours_q.pdf}
    \subcaption{Example of correspondence between query graph and density estimators. Density estimators for closed in-neighborhood subschemas that overlap with query graph and are not redundant are used. \label{fig:g_prop_q}}
\end{minipage}
\caption{Method based on subschemas partitioned per closed in-neighborhood \label{fig:g_prop}}
\normalsize
\end{figure}

In the join cardinality estimation, it is challenging to achieve efficient execution and high inference accuracy when the schema size is increased.
In Scardina, we attempt to solve these issues by employing an approach using multiple small density estimators, as shown in \cref{fig:g_prop}.
There are two main advantages to carefully partitioning the range covered by each density estimator into small parts.
The first is that it requires a more compact model with fewer parameters and can be trained in parallel.
The second is that it can improve estimation accuracy for small query graphs, which are often required for practical applications such as query optimizers.
Schema partitioning using only the structure of the schema graph avoids the additional cost of partitioning that is incurred with DeepDB.
In general, density estimators for each subschema lose information across tables, resulting in poor estimation quality.
Scardina, therefore, achieves better approximation by considering the correlation among density estimators through overlapping tables.

The main flow of Scardina is as follows for each of the preprocessing phase, training phase, and inference phase:
\begin{enumerate}
    \item Preprocessing Phase
    \begin{enumerate}
        \item Partition a global schema based on its structure into subschemas.
    \end{enumerate}
    \item Training Phase
    \begin{enumerate}
        \item Create full outer joined tables for each subschema.
        \item Tranin density estimators for each full outer joined table with join-sampled tuples.
    \end{enumerate}
    \item Inference Phase
    \begin{enumerate}
        \item Select density estimators required for answering a query.
        \item Set the size of the starting table as the initial value of the estimation
        \item Infer the probability of satisfying predicates with a density estimator to update the estimated value.
        \item When crossing subschemas, Fanouts are inferred to update the estimated value for upscaling by joins, and subsequent inferences use samples from the preceding common table as inference conditions.
        \item Traversing all the subschemas required for the query yields an estimated cardinality.
    \end{enumerate}
\end{enumerate}

\subsection{Partitioning Global Schema into Closed In-neighborhood Subschemas \label{ssec:prop_partition}}
This section describes the procedure for determining the subschemas to be estimated for each density estimator used in Scardina.
First, we explain the case where the global schema graph $G$ is a directed simple graph, i.e., without multiple edges (See \cref{sec:multiedges} for details on the case of handling multigraphs).
Scardina performs schema partitioning per closed in-neighborhoods of the global schema graph vertices.
Taking the global schema graph in \cref{fig:schema_g} as an example, partitioned as in \cref{fig:g_prop_s}.
Focusing on each vertex, there are three vertices with an incoming degree greater than or equal to 1, $S, U, W$, and so the schema is partitioned into three closed in-neighborhood subschemas centered on $S, U, W$.

We describe the properties of the closed in-neighborhood subschema by Scardina.
As noted in \cref{ssec:notation}, directed edges of schema graphs represent one-to-many relationships.
Therefore, all subgraphs $G[N^{-}_G[v]]$ induced by a vertex set $N^{-}_G[v]$ that is a closed in-neighborhood are subschemas that have only many-to-one relationships around a table $v$.
This means that a center table is never scaled in a full outer join of closed in-neighborhood tables, and only open in-neighborhood tables are scaled.
As a result, downscaling in the subschema, which is necessary with universal relation-based and correlation-based partitioned methods, is no longer necessary during inference, leading to improved inference accuracy (See \cref{ssec:prop_query} for details on downscaling).
Next, we discuss how the closed in-neighborhood subschemas can handle various types of relationships that may appear in databases.
There are three main types of relationships: one-to-one, one-to-many, and many-to-many.
Of the three types of relationships, one-to-one is simply a vertical partition of a single table, which can be treated as a single table without scaling by vertically merging the tables.
Therefore, no special consideration is required.
Second, one-to-many is represented, for example, when modeling publishers and books.
While a publisher can publish multiple books, it is assumed that a book is never published by more than one publisher.
In this case, scaling occurs with the number of books, which needs to be captured by a density estimator.
Since each one-to-many relationship represented by a directed edge is always contained in one of the closed in-neighborhood subschemas, the density estimator of the corresponding subschema can appropriately capture the distribution.
Finally, many-to-many is represented, for example, when modeling authors and books.
Consider that an author may write several books, and a book may be written by several authors.
In this case, scaling occurs depending on the number of both authors and books, and this needs to be captured.
Given the normalization of this relationship, we can represent authors and writings as one-to-many, and writings and books as many-to-one, so that we can represent them as two one-to-many without loss of information.
Since each closed in-neighborhood subschema contains all the many-to-one relationships from the central table, resulting in a normalized many-to-many relationship that can be handled in a single density estimator.
This property is not common, for example, when using closed out-neighborhood partitioning, it spans many-to-many relationships into multiple subschemas.
Therefore, it can be said that density estimation based on closed in-neighborhood partitioning is superior for handling relationships between two tables in practice.

\subsection{Training Estimators for each Closed In-neighborhood Subschema \label{ssec:prop_model}}
The following part describes the training of density estimators based on obtained closed in-neighborhood subschemas.
Similar to the universal relation-based method (\cref{ssec:relwork_univrel}) and the correlation-based partitioning method (\cref{ssec:relwork_corrpart}), Scardina uses a single estimator to handle queries on a subset of tables contained in the subschema, as well as inner and outer joins, so that Scardina employs a full outer joined table of all tables contained in the subschema.
The additional virtual attribute, TableFlagFanout, is handled in a similar manner.
However, instead of Fanouts between tables in a closed in-neighborhood subschema, Fanouts to adjacent tables outside the subschema are required\footnote{More precisely, Fanouts within a subschema are required only when estimating the cardinality of a single table with zero in-degree. Since this paper focuses on join cardinality, we omit such cases in this paper.}.
The density estimator should be able to be conditioned from outside based on specific estimation results.
If conditioning is not possible, the estimation accuracy will deteriorate in cases where query graphs traverse multiple closed in-neighborhood subschemas, as estimation is performed independently for each subschema.
For example, Sum-Product Network~\cite{sumproductnetworks} used in DeepDB is a kind of density estimator, but it is not suitable for Scardina because the estimation target is closed.

Here, we use Denoising Autoencoders as density estimators, which have been confirmed to perform consistently well in AutoEncoderUR\cite{nar_cardest} and can handle conditioning.
Similar to AutoEncoderUR, training is performed by randomly selecting attributes from among all attributes as noise, masking the selected attributes, and using them as input.
Parameters are updated by computing cross-entropy loss on the output of the masked attributes.
This training makes density estimators capable of estimating distributions of any attributes conditional on any attributes.

The training data is full outer joined tuples of all the tables in closed in-neighborhood subschemas.
Although it is possible to simply use all tuples from joined tables to train, it would take too much time when the number of data is quite large, while training would not progress well when the number of data is too small.
To mitigate the effect of the number of data, we use join sampling.
In order to obtain i.i.d. samples from an original joint distribution, we use Exact Weight Algorithm~\cite{joinsampling}, which is also used in NeuroCard.
This sampling allows for sufficient scaling to the data size without compromising Scardina's scalability.
Although density estimators are trained for each closed in-neighborhood subschema, each density estimator can be trained independently, since join samples used by each estimator are unique to each estimator.
In other words, density estimators can be trained in parallel for each subschema, allowing for a short start-up time of Scardina.

\subsection{Querying over Multiple Estimators \label{ssec:prop_query}}
We describe a join cardinality estimation procedure using multiple density estimators of closed in-neighborhood subschemas.
In order to perform inference across multiple density estimators, Scardina regards a set of closed in-neighborhood subschemas as a hypergraph and performs inference by traversing hyperedges based on a query graph.
Intuitively, it is like preparing a starting table and emulating the iterative process of predicate filtering and joining with subsequent tables.

First, we explain the construction of a hypergraph from a set of closed in-neighborhood subschemas.
The vertex set of each closed in-neighborhood is a hyperedge $\mce \in \mcE$ and a hypergraph $H$ is constructed by combining a set of closed in-neighborhoods as hyperedges $\mcE := \{ N^{-}_{G}[v] \mid v \in V_{GLB}\}$ and the vertex set contained in the global schema graph $V_{GLB}$.
In this case, the following \cref{def:connected} holds.
\begin{lem} \label{def:connected}
If a directed graph $G$ is connected when viewed as an undirected graph, a hypergraph $H$ with $\mcE := \{ N^{-}_{G}[v] \mid v \in V\}$ as hyperedges is also connected.
\begin{proof}
Let $H' := (V, \mcE')$ be $G$ viewed as a hypergraph with hyperedges $\mcE' := \{\{u, v\} \mid (u, v, c) \in E \}$.
Then $H'$ is obviously connected.
By the definition of the closed in-neighborhood, let closed in-neighborhoods be hyperedges, then all adjacent vertices are contained in the same hyperedge.
Therefore, $\FA\mce \in \mcE, \EX\mce' \in \mcE' \text{ s.t. } \mce' \subseteq \mce \land \FA \mce' \in \mcE, \EX \mce \in \mcE \text{ s.t. } \mce' \subseteq \mce$.
That is, $H$ is also connected since it contains the connectivity of $H'$.
\end{proof}
\end{lem}

Next, we describe the inference procedure based on the constructed hypergraph and a query graph.
Scardina's inference procedure is shown in \cref{alg:query}.
First, since a query does not always access all tables, its query graph $G_Q$ is used to narrow down to necessary closed in-neighborhood subschemas, that is, to construct the corresponding subhypergraph $H_Q$ (Line~\ref{l:subhe}~-~\ref{l:subhg}).
Create a tree to traverse the hyperedge $\mcE_Q$ of the obtained subhypergraph $H_Q$ (Line~\ref{l:totree}).
Randomly select a starting point $\mce_1$ from the hyperedges $\mcE_Q$ and set the size of its closed in-neighborhood joined table $|J_{e_1}|$ as the initial value of cardinality (Line~\ref{l:source}).
And then, traverse the hyperedges in breadth-first order (Line~\ref{l:for_bfs_s}).
For each hyperedge, we first enumerate attributes that need to be estimated in addition to the predicates (Line~\ref{l:e_attrs}~-~\ref{l:fanout2}).
Specifically, $\bA_{common}$, the attributes of the vertices that are common with subsequent hyperedges $\mcf$ and Fanouts $\bF$ that are required when evaluating joins with subsequent hyperedges.
These variables and the probabilities of satisfying the predicates are inferred by the density estimator $\mcM_{J_{\mce}}$ of the closed in-neighborhood joined table $J_{\mce}$ corresponding to the hyperedge $\mce$ (Line~\ref{l:est}).
If there is a preceding hyperedge, it uses common table attributes $common\_samples$ as input.
Inference is performed using AutoEncoderUR~\cite{nar_cardest}'s inference algorithm with the addition of input samples $common\_samples$, additional inference of common table attributes $\bA_{common}$, and Fanouts $\bF$ (\cref{alg:est}).
$P_{J_{\mce}}(R_Q)$, the resulting probability of satisfying predicates in $J_{\mce}$, and Fanouts $fanouts$ are multiplied to the intermediate cardinality (Line~\ref{l:updatecard1}~-~\ref{l:updatecard2}).
Performing above operations on all hyperedges $\mcE_Q$ yields the estimated cardinality $\hat{\mcC(Q)}$ for the query $Q$ (Line~\ref{l:returncard}).

\begin{table}[ht]
\centering
\caption{Function definitions used in \cref{alg:query} \label{tbl:func_query}}
\begin{tabularx}{\linewidth}{l|X} \hline
Function Name & Definition \\ \hline \hline
$\textsc{Random}(\bm{x})$ & Select one element randomly from $\bm{x}$. \\ \hline
$\textsc{ToTree}(H, G)$ & Construct tree from edge set $\mcE$ of hypergraph $H$, covering directed graph $G$. By \cref{def:connected}, connected tree should be obtained. \\ \hline
$\textsc{BFS}(T, s)$ & Enumerats edges of tree $T$ in breadth-first order, with $s$ as starting point, in the form of tuples consisting of vertex and set of following vertices. \\ \hline
$\textsc{AttributesIn}(V)$ & Retrieve flattened attribute set of vertex set $V$. \\ \hline
\rowsl{$\textsc{EdgeToFanout}$}{$\textsc{Attribute}(u, v, c)$} & Retrieve Fanout attributes from table $u$ to $v$ with constraint $c$. \\ \hline
\end{tabularx}
\end{table}

\begin{algorithm}[ht]
\caption{Join cardinality estimation using closed in-neighborhood density estimators \label{alg:query}}
\small
\begin{algorithmic}[1]
\INPUT{
Global schema graph $G_{GLB}:=(V_{GLB},V_{GLB})$,
Hypergraph comprising closed in-neighborhood subschemas $H:=(V_{GLB}, \mcE)$,
Acyclic query $Q:=(R_Q, G_Q)$, $G_Q \subseteq G_G$, $G_Q$ is simple tree
}
\OUTPUT{Estimated cardinality of $Q$ $\hat{\mathcal{C}(Q)}$}
\State{$\mcE_Q \gets \{\mce \in \mcE \mid \mce \setminus V_Q \ne \emptyset\}$ \label{l:subhe}} %
\State{$H_Q := (V_Q, \mcE_Q)$} \label{l:subhg}
\State{$\mce_1 \gets \textsc{Random}(\mcE_Q)$} \label{l:source}
\State{$common\_samples \gets \{\}$}
\State{$\hat{\mcC} \gets |J_{\mce_1}|$} \label{l:initcard} \Comment{Number of rows of root table as initial cardinality}
\State{$T_{\mcE_Q} \gets \textsc{ToTree}(H_Q, G_Q)$} \label{l:totree} \Comment{Restructure hyperedges to tree w.r.t. tree $G_Q$}
\State{$successor\_list \gets BFS(T_{\mcE_Q}, \mce_1)$} \label{l:bfs} \Comment{Arrange hyperedges in BFS order}
\ForEach{$\mce, \mbcf \in successor\_list$} \label{l:for_bfs_s} \Comment{$\mce$ and $\mbcf$ are current hyperedge and successors, respectively}
  \State{$\bA \gets \textsc{AttributesIn}(\mce)$} \label{l:e_attrs}
  \State{$\bA_{common} \gets
    \bigcup_{\mcf \in \mbcf}(\bA \cap \textsc{AttributesIn}(\mcf))$} \label{l:commonattrs}
  \State{$fanout\_edges \gets
    \bigcup_{\mcf \in \mbcf}(\{(u,v,c) \in E_{G_Q[\mcf]} \mid {}^{\exists}u \in (\mce \cap \mcf)\})$} \label{l:fanout1}
  \State{$\bF \gets \textsc{EdgeToFanoutAttribute}(fanout\_edges)$} \label{l:fanout2}
  \State{$\hat{P}_{J_{\mce}}(R_Q), \hat{fanouts}, \hat{samples} \gets \textsc{Estimate\_N}(\mcM_{J_{\mce}}, R_Q, \bA, \bF, \bA_{common}, common\_samples)$} \label{l:est}
  \State{$\hat{\mcC} \gets \hat{\mcC} \times \hat{P}_{J_{\mce}}(R_Q)$} \label{l:updatecard1}
  \State{$\hat{\mcC} \gets \hat{\mcC} \times \prod_{\hat{fanout} \in \hat{fanouts}}\hat{fanout}$} \label{l:updatecard2}
  \State{$common\_samples.add(\hat{samples})$}
\EndFor \label{l:for_bfs_e}

\State {$\textbf{return} \ \hat{\mcC} \text{ as } \hat{\mcC(Q)}$} \label{l:returncard}
\end{algorithmic}
\normalsize
\end{algorithm}

\begin{table}[ht]
\centering
\caption{Function definitions used in \cref{alg:est} \label{tbl:func_est}}
\begin{tabularx}{\linewidth}{l|X} \hline
Function Name & Definition \\ \hline \hline
$\textsc{DrawSample}(dist_A)$ &
    Draw samples from $dist_A$, the distribution of attribute $A$, using $dist_A$ as weights. \\ \hline
$\textsc{Encode}(a)$ &
    Embed an element $a$ of an attribute $A$.
    One-hot vectors or Entity Embeddings~\cite{entityembeddings} can be used. \\ \hline
$\textsc{SortBy}(\bA, key\_func)$ &
    Sort attributes $\bA$ in ascending order based on $key\_func$. \\ \hline
\end{tabularx}
\end{table}

\begin{algorithm}[ht]
\caption{Using the density estimator to infer the selectivity and Fanouts and to obtain samples  \label{alg:est}}
\small
\begin{algorithmic}[1]
\INPUT{
Density Estimator $\mcM$,
Predicate ranges $R_Q$,
Whole attributes $\bA$,
Fanout attributes $\bF$,
Attributes to be additionally sampled $\bA_{sample}$,
Samples $samples$
}
\OUTPUT{Estimated selectivity $\hat{\mcS}$, Fanouts $\hat{fanouts}$, Samples w/ some updates $\hat{samples}$}

\Procedure{Estimate\_N}{} \Comment{Sample size is $N$}
\Procedure{Estimate}{$\mathcal{M}, R_Q, \bA_Q, \bF, \bA_{sample}, sample$}
  \State {Initialize $inputs$ with $sample$}
  \State $\hat{prob} \gets 1.0 $
  \ForEach{$A \in \bA_Q$} \Comment{Estimate probabilities of predicates}
    \State {$\hat{dist_A} \gets \mcM(inputs)$} \Comment{Forward}
    \State {$\hat{dist'}_A \gets \{ \hat{dist_a} * (a \in R_Q(A)) \mid a \in A \}$} \Comment{Filter distribution by $R_Q$}
    \State {$\hat{prob} \gets \hat{prob} * \sum_{a \in A}\hat{dist'_a}$}
    \State {$\hat{a} \gets \textsc{DrawSample}(\hat{dist'_A})$}
    \State {$inputs[A] \gets \textsc{Encode}(\hat{a})$}
  \EndFor
  \ForEach{$A \in \bF$} \Comment{Estimate fanouts}
    \State{$\hat{dist_A} \gets \mcM(inputs)$}
    \State{$\hat{a} \gets \textsc{DrawSample}(\hat{dist_A})$}
    \State{$\hat{fanout}[A] \gets \hat{a}$}
    \State{$\hat{sample}[A] \gets \hat{a}$}
    \State{$inputs[A] \gets \textsc{Encode}(\hat{a})$}
  \EndFor
  \ForEach{$A \in \bA_{sample}$} \Comment{Draw samples for subsequent estimation}
    \State{$\hat{dist_A} \gets \mcM(inputs)$}
    \State{$\hat{a} \gets \textsc{DrawSample}(\hat{dist_A})$}
    \State{$\hat{sample}[A] \gets \hat{a}$}
    \State{$inputs[A] \gets \textsc{Encode}(\hat{a})$}
  \EndFor
  \State {$\textbf{return} \ \hat{prob}, \hat{fanout}, \hat{sample}$}
\EndProcedure
\State
\State {$\bA_Q \gets \{ A \in \bA \mid |R_Q(A)| < \text{dom}(A) \}$} \label{l:filterattr} \Comment{Filter attributes by predicates}
\State {$\bA_Q \gets \textsc{SortBy}(\bA_Q, key: A \rightarrow |R_Q(A)|)$} \label{l:sortattr} \Comment{Sort attributes by domain size}
\ForEach{$i \in \{1, \dots, N\}$} \Comment{Batched in practice}
  \State {$\hat{probs}[i], \hat{fanouts}[i], \hat{samples}[i] \gets \textsc{Estimate}(\mathcal{M}, R_Q, \bA_Q, \bF, \bA_{sample}, \hat{samples})$}
\EndFor
\State {$\hat{\mathcal{S}} \gets \textsc{mean}(\hat{probs})$}
\State {$\textbf{return} \ \hat{\mathcal{S}}, \hat{fanouts}, \hat{samples}$}
\EndProcedure
\end{algorithmic}
\normalsize
\end{algorithm}

The following is a concrete explanation of the inference procedure by Scardina, using \cref{fig:g_prop} as an example.
Let a starting point $\mce_1$ be $\{S, T\}$, the initial value of cardinality is $|S \Join T|$.
First, infer the probability of satisfying the query condition $Q$ under $S \Join T$ and the fanout $F_{T.id=U.t\_id}$ when $U$ is joined to $S \Join T$ with $\mcM_{S \Join T}$.
Next, infer the probability of satisfying the query condition $Q$ under $T \Join U \Join V$ with $\mcM_{T \Join U \Join V}$ using samples of the attributes of $T$ obtained from $\mcM_{S \Join T}$ as input.
Finally, the final estimated join cardinality is obtained by computing the product of the initial value, the inferred probabilities, and Fanouts.

This estimation strategy has the following two distinctive features compared to existing methods:
The first is that samples of attributes from common tables are taken as input between hyperedges, so that inference is performed under the conditions of the preceding attributes, even though they are limited to common tables.
This is always possible because Denoising Autoencoder, which is used as a density estimator, has the property of being able to infer arbitrary conditions, and because of connection as shown in \cref{def:connected}.
This allows inference across multiple density estimators without assuming complete independence and is expected to improve overall inference accuracy.
The second is that downscaling is not required even if a hyperedge $\mce \in \mcE_Q$ contains tables that are not necessary for a query (${}^{\exists}v \in \mce \land v \notin V_Q$).
This is because $\mce$ is based on a closed in-neighborhood $N^{-}_G[v]$, and unnecessary tables are always many-to-one from the viewpoint of $v$.
Assuming $v$ as the center table, possible unnecessary tables are limited to $N^{-}_G(v)$\footnote{$\because$ Except when $|V_Q|$ is 1, if $v$ is unnecessary, it is because $\mathcal{e}$ itself is redundant and unnecessary.}.
This also reduces additional inference and contributes to the overall improvement of inference accuracy.

Finally, we formulate Scardina's join cardinality estimation method.
Let $\bJ_Q$ be a set of closed in-neighborhood joined tables sufficient for estimating the join cardinality of a query $Q$, and let $J_1$ be the first subschema taken out.
The inference by Scardina can be expressed as \cref{eq:prop} using \cref{eq:def_prob}.
Here, the expectation represented by $\bbE_{t_j \sim \mcM_{J_j} \mid J_{j}.\bm{A} = t_{<j}.\bm{A}}$ denotes the use of the attribute set $\bA$ of the common table $T \in (V_{J_j} \cap V_{J_{<j}})$ among the samples $t_{<j}$ of $J_{<j}$ preceding $J_j$ as a condition for inference.
Compared to the universal relation-based method (\cref{eq:neurocard}), it is inferior in terms of rigor due to the existence of an approximation for each expectation, but compared to the correlation-based partitioning method (\cref{eq:deepdb}), it tends to be more accurate in terms of approximation due to the fact that each expectation is no longer independent.
Also, in contrast to both \cref{eq:neurocard} and \cref{eq:deepdb}, downscaling by Fanouts is not necessary, which is also likely to contribute to enhanced accuracy.
More detailed rigor comparisons by partitioning strategy are shown in \cref{sec:diffpartition}.

\begin{table*}[bt]\begin{tabular}{c}
$\begin{aligned}
\mathcal{C}_{INP}(Q)
&=
  |J_1|
  \cdot
    \prod_{i=1}^{N}
        \underset{t_i \sim \mcM_{J_{i}}}{\bbE}
        \Biggl\lbrack
        \mathbb{I}(\bA, Q_{J_i}, t_i)
        \cdot 
        \prod_{
            \substack{(T_1, T_2, (A_1, A_2)) \in E_Q \\
    \land T_1 \in V_{J_i} \land T_2 \notin V_{J_i} \land T_2 \in V_{J_{>i}}}
        }{
            t_i[F_{(T_1, T_2, (A_1, A_2))}]
        }
        \Biggm|
        \Big\langle \FA{j} \in [1,i-1] \land \FA{A} \in (J_i.\bA \cap J_j.\bA) \Big\rangle \land J_i.A = t_j[A]
        \Biggr\rbrack
  \eqlabel{eq:prop}
\end{aligned}$
\end{tabular}\end{table*}

\section{Evaluation \label{sec:eval}}
\memo{WIP}
In this section, we evaluate the join cardinality estimation methods.
First, we describe the experimental setup in \cref{ssec:setup}, then evaluate cardinality estimation separately in \cref{ssec:exp_cardest}, and evaluate cardinality estimation when used for the query optimizer in \cref{ssec:exp_qo}.

\subsection{Experimental Setup \label{ssec:setup}}
\subsubsection{Benchmarks \label{sssec:benchmark}}
We describe the benchmarks, combinations of dataset and workload, used in the experiments.

\xparagraph{Join Order Benchmark}
Join Order Benchmark~\cite{cardest} utilizes the IMDb\footnote{url{https://www.imdb.com}} dataset, a real-world dataset of actors, entertainment, and video games.
It consists of 16 tables with up to 36M rows and 17 attributes.
Two workloads are used for evaluation: JOB-light~\cite{learnedcardinalities} and JOB-m~\cite{neurocard}.
JOB-light consists of 70 queries with multiple equality and range conditions and inner joins of up to 5 tables.
JOB-m consists of 113 queries with multiple equality and range conditions as well as IN clauses and LIKE operators, and performs inner joins on up to 11 tables.

\subsubsection{Metrics \label{sssec:metrics}}
\xparagraph{\textsf{Q-Error}}
For the evaluation metric of estimation accuracy, we use \textsf{Q-Error}~\cite{cardest}, represented by \cref{eq:qerr}, where $\mcC$ is the true cardinality and $\hat{\mcC}$ is the estimated cardinality. This is a dimensionless value that indicates how many times the estimated cardinality is away from the true value and is always at least $1$, with smaller values being better.
\begin{align}
\textsf{Q-Error} := \frac{\text{max}(\mcC, \hat{\mcC})}{\text{min}(\mcC, \hat{\mcC})} \label{eq:qerr}
\end{align}

\xparagraph{\textsf{P-Error}}
Furthermore, we use \textsf{P-Error}~\cite{statsperr} as the metric corresponding to the query execution time.
For this metric, we first use PostgreSQL's query optimizer to create execution plans using the true cardinality $\mcC$ and the estimated cardinality $\hat{mcC}$, respectively.
Since the purpose of this process is to create execution plans, the cardinality estimation for multiple parts of the target query is used, not for the entire query.
Therefore, the cardinality estimation accuracy depends more on the estimation accuracy of small queries that are a subset of entire queries rather than on larger parts of or entire queries.
Next, the PostgreSQL query optimizer is used again to estimate the execution costs of the two created execution plans.
Finally, the ratio of the estimated execution costs obtained is \textsf{P-Error}.
In most cases, it is greater than or equal to $1$, indicating that the smaller the better.
Let $P(\mcC)$ denote the execution plan by the query optimizer using cardinality $\mcC$ and $PCEst(P)$ denote the estimated execution cost of the execution plan $P$, then \textsf{P-Error} can be expressed as \cref{eq:perr}.
\begin{align}
\textsf{P-Error} := \frac{PCEst(P(\hat{\mcC}))}{PCEst(P(\mcC))} \label{eq:perr}
\end{align}
True cardinality is used for cardinality estimation other than execution plan creation.
This metric is unique in that it allows comparisons that are correlated to query execution time comparisons without the need for time-consuming actual query execution.
Depending on the query optimizer, there is a slight possibility that the estimated cost of an execution plan using true cardinality exceeds that using estimated cardinality: $\textsf{P-Error}<1$.
However, since $PCEst(P(\mcC))$ is consistent across comparison methods, it is considered to be a reasonable metric for comparison.

In practice, cardinality estimation is performed by dividing the target query into subsets per join in advance, and the estimated value set is given to the PostgreSQL optimizer via CEB~\cite{cebgithub} and pg\_hint\_plan~\cite{pghintplan} to obtain an execution plan $P(\hat{\mcC})$.
Then, the PostgreSQL optimizer estimates the execution cost $PCEst(P(\hat{\mcC})$ of the obtained $P(\hat{\mcC})$.
Similarly, the execution cost $PCEst(P(\mcC))$ when using true cardinality is also estimated, and finally, \textsf{P-Error} is derived.

\subsubsection{Methods \label{sssec:methods}}
\textsf{Scardina} uses the Denoising Autoencoder-based model proposed in AutoEncoderUR~\cite{nar_cardest} as a density estimator.
The implementation of the Denoising Autoencoder uses Multi-Layer Perceptron.

For the comparison methods, we use (1)\textsf{PostgreSQL}(version 11.8), which uses desity estimation by statistics, (2)\textsf{AutoEncoderUR}~\cite{nar_cardest} and (3)\textsf{NeuroCard}~\cite{neurocard,neurocardgithub}, which use density estimation by a distribution of the universal relation, and (4)\textsf{DeepDB}~\cite{deepdb,deepdbgithub}, which use density estimation by distributions of correlation-based partitioned subschemas.
The hyperparameters of the existing methods use the values reported in the respective papers or publicly available source codes ~\cite{neurocard,deepdb,neurocardgithub,deepdbgithub}.

\subsubsection{Environment \label{sssec:env}}
We evaluate all methods in the same runtime environment with 16 CPUs (up to 2.5 GHz), 64 GB of memory, and Nvidia Tesla T4 GPUs.
Only \textsf{Scardina} and \textsf{NeuroCard} can be accelerated by GPUs.

\subsection{Experiment 1: Accuracy of Cardinality Estimation \label{ssec:exp_cardest}}
\xparagraph{Evaluation guideline and motivation}
We evaluate the performance of join cardinality estimation standalone.
We evaluate each method based on its \textsf{Q-Error}(Median, 90th percentile, 95th percentile, 99th percentile, and maximum value), average response time, and training time, and confirm the usefulness of \textsf{Scardina}.

\xparagraph{Results}
The results of the evaluation using \textsf{Q-Error} for JOB-light and JOB-m are shown in \cref{tbl:qerr_jobl} and \cref{tbl:qerr_jobm}, respectively.
JOB-light always showed the best estimation accuracy for \textsf{Scardina}.
Since JOB-light only accesses a maximum of 5 out of 16 tables, building a density estimator for each closed-entry neighborhood subschema seems to work well.
Similarly, for the average response time, the number of parameters can be relatively reduced because the density estimator is built for each subschema, which contributes to faster inference.
On the other hand, existing methods such as \textsf{NeuroCard} and \textsf{AutoEncoderUR} use a large density estimator for the universal relation and require a large number of downscaling during inference, resulting in low estimation accuracy and low response performance.
\textsf{DeepDB} does not work because it requires more than 100 GB of memory for the correlation calculation for schema partitioning.
The same trend would be expected for datasets of the same or larger size as the IMDb dataset used in this experiment, indicating that there is scalability concern.
In terms of response time alone, the fastest response time is achieved by \textsf{PostgreSQL}, which uses histograms, but this is a trade-off with estimation accuracy.

In JOB-m, \textsf{Scardina} outperforms \textsf{PostgreSQL}, but is not as good as {textsf{NeuroCard}}.
In contrast to the JOB-light case, JOB-m workload accesses up to 11 tables, which is considered to be an advantage for universal relations.
Compared to the JOB-light case, the response times for \textsf{Scardina} and \textsf{NeuroCard} increased significantly, which is likely a common implementation issue related to the IN clause and LIKE operator in the JOB-m query.
Both \textsf{Scardina} and \textsf{NeuroCard} take a lot of time in the process of performing those conditionals for the whole domain.
We believe that indexing and efficient evaluation of conditions can improve the performance.

Looking at the training time of the density estimators used in common by the two workloads, we see that \textsf{Scardina} has the shortest training time.
This is the result of the fact that the method can be trained in parallel and that the number of parameters for each density estimator is small.
Compared to methods using the universal relation, especially when accessing a large number of tables, the estimation accuracy can be inferior due to the approximation across density estimators, as in the JOB-m results.
However, for larger schemas, considering the training time as well, \textsf{Scardina} is more suitable for practical use.
In addition, the fact that the density estimators are independent has other advantages besides the parallel training.
In the future, when we assume an environment where data changes, we may update the density estimator to keep up with the data changes.
In this case, Scardina can be used to update only the density estimators that contain the updated tables, rather than the entire large density estimator.

\footnotetextx{cnt_samemodel}{Since both are evaluated on the same model trained on the same dataset, the training times are identical.}
\footnotetextx{cnt_pgexplain2}{The response times are reported from EXPLAIN statements and are for reference only. The time required for cardinality estimation only is shorter than the times shown.}

\begin{table*}[!tb]
\centering
\caption{Percentiles of \textsf{Q-Error}, average response time (ms), and training time (min) for each method in JOB-light\protect\footnotemarkx{cnt_samemodel}} \label{tbl:qerr_jobl}
\scalebox{0.9}{
\begin{tabular}{l||r|r|r|r|r||r|r} \hline
Method & Median & 90\%th & 95\%th & 99\%th & Max & \scalebox{0.6}[1]{Response Time (ms)} & \scalebox{0.6}[1]{Training Time (min) } \\ \hline \hline
\SB{0.85}{\textbf{Ours}} & 1.60 & 4.19 & 6.87 & 19.7 & 33.0 & 22.8 & 66.3\tablefootnote{Assuming sufficient computational resources are available.} \\
\SB{0.85}{\textsf{AutoEncoderUR}} & 1.68 & 5.66 & 22.1 & 33.6 & 34.8 & 50.4 & 128 \\
\textsf{NeuroCard} & $1.79$ & $9.00$ & $19.5$ & $36.5$ & $43.2$ & 160 & 391 \\
\textsf{DeepDB} & \multicolumn{7}{c}{\textit{---Out of memory to train (100GB+)---}} \\
\textsf{PostgreSQL}\footnotemarkx{cnt_pgexplain2} & $7.44$ & $163$ & $1.1 \cdot 10^3$ & $2.8 \cdot 10^3$ & $3.5 \cdot 10^3$ & 3.44 & --- \\ \hline
\end{tabular}
}
\end{table*}

\begin{table*}[!tb]
\centering
\caption{Percentiles of \textsf{Q-Error} and \textsf{wQ-Error}, average response time (ms), and training time (min) for each method in JOB-m\protect\footnotemarkx{cnt_samemodel}} \label{tbl:qerr_jobm}
\scalebox{0.85}[0.9]{
\begin{tabular}{l||r|r|r|r|r|r||r|r} \hline
Method & Median & 90\%th & 95\%th & 99\%th & Max & \scalebox{0.6}[1]{Response Time (ms)} & \scalebox{0.6}[1]{Training Time(min) } \\ \hline \hline
\SB{0.85}{\textbf{Ours}} & 3.72 & 378 & $1.4 \times 10^3$ & $1.6 \times 10^4$ & $8.7 \times 10^4$ & $1.7 \times 10^3$ & 66.3 \\
\textsf{NeuroCard} & 2.74 & 160 & 559 & $4.8 \cdot 10^3$ & $1.7 \cdot 10^4$ & 846 & 391 \\
\textsf{DeepDB} & \multicolumn{7}{c}{\textit{---Out of memory to train (100GB+)---}} \\
\textsf{PostgreSQL}\footnotemarkx{cnt_pgexplain2} & 160 & $5.8 \cdot 10^3$ & $1.3 \cdot 10^4$ & $8.7 \cdot 10^4$ & $1.0 \cdot 10^5$ & 6.95 & --- \\ \hline
\end{tabular}
}
\end{table*}

\subsection{Experiment 2: Impact on Query Optimizer \label{ssec:exp_qo}}
\xparagraph{Evaluation guideline and motivation}
As one of the applications of the join cardinality estimation, we evaluate \textsf{Scardina} for use in the query optimizer.
We evaluate each method in terms of its \textsf{P-Error} (Median, 90th percentile, 95th percentile, 99th percentile, and maximum value) and confirm that the \textsf{Scardina} contributes to produce better execution plans.

\xparagraph{Results}
The results of the evaluation of JOB-light and JOB-m by \textsf{P-Error} are shown in \cref{tbl:perr_jobl} and \cref{tbl:perr_jobm}, respectively.
First, on JOB-light, \textsf{Scardina} has the best results.
This means that query processing through the query optimizer can be accelerated.
On the other hand, on JOB-m, we observe the performance of \textsf{Scardina} is almost the same or slightly inferior to the existing methods.
The reason for this is that, as can be seen in \textsf{Q-Error} of \cref{tbl:qerr_jobm}, when a large number of inferences across partitioned subschemas are required, the effect of correlations that cannot be propageted between density estimators other than overlapping tables increases.
This is likely to lead to accuracy degradation.
However, this degradation in estimation capability is a trade-off between the time required for individual inference and the training time, as shown in \cref{ssec:exp_cardest}.
Scalability as provided by \textsf{Scardina} is vital, especially in environments with large schemas, since it is essential to avoid the scenario in which the system does not work in the first place.
In the future, we may evaluate \textsf{Scardina} on benchmarks with larger schemas.

\begin{table}[!htb]
  \centering
  \caption{Percentiles of \textsf{P-Error} on JOB-light \label{tbl:perr_jobl}}
  \begin{tabular}{l||r|r|r|r|r} \hline
  Method & Median & 90\%th & 95\%th & 99\%th & Max \\ \hline \hline
  \SB{0.85}{\textbf{Ours}} & $1.00$ & $1.17$ & $1.32$ & $1.97$ & $2.26$ \\
  \SB{0.85}{\textsf{AutoEncoderUR}} & 1.00 & 1.40 & 1.80 & 2.34 & 2.41 \\
  \textsf{NeuroCard} & 1.01 & 1.56 & 2.27 & 2.53 & 4.11 \\
  \textsf{DeepDB} & \multicolumn{5}{c}{\textit{---Out of memory to train (100GB+)---}} \\
  \textsf{PostgreSQL} & $1.00$ & $1.24$ & $1.34$ & $2.01$ & $2.63$ \\ \hline
  \end{tabular}
\end{table}

\begin{table}[!htb]
  \centering
  \caption{Percentiles of \textsf{P-Error} on JOB-m \label{tbl:perr_jobm}}
  \begin{tabular}{l||r|r|r|r|r} \hline
  Method & Median & 90\%th & 95\%th & 99\%th & Max \\ \hline \hline
  \SB{0.85}{\textbf{Ours}} & $1.00$ & $3.12$ & $7.53$ & $37.8$ & $50.4$ \\
  \textsf{NeuroCard} & 1.02 & 2.33 & 2.94 & 15.7 & 50.4 \\
  \textsf{DeepDB} & \multicolumn{5}{c}{\textit{---Out of memory to train (100GB+)---}} \\
  \textsf{PostgreSQL} & $1.03$ & $3.15$ & $6.27$ & $36.9$ & $50.3$ \\ \hline
  \end{tabular}
\end{table}

\section{Conclusion \label{sec:conc}}
In this paper, we proposed \textbf{Scardina}, a scalable join cardinality estimation method using multiple density estimators.
The main features of Scardina can be summarized in three points:
The first is scalability with respect to schema size.
We perform a simple partitioning of a database schema into subschemas based on the schema graph structure and train density estimators for each subschema.
Since each density estimator can be compact and independent, training becomes easier and it enables training in parallel.
The second is scalability with respect to data size.
By properly sampling training data and using them, it is possible to flexibly handle small to large-scale data.
The third is high estimation accuracy.
Small density estimators are easy-to-fit to queries, and it avoids unnecessary inferences and inaccurate assumptions that are inconsistent with the data.
This is especially useful when used by query optimizers.

We confirmed that Scardina scales with respect to the schema size and is effective for query optimizers.

\begin{acks}
This paper is based on results obtained from a project, JPNP16007, commissioned by the New Energy and Industrial Technology Development Organization (NEDO).
\end{acks}

\bibliographystyle{ACM-Reference-Format}
\bibliography{ref}

\clearpage

\appendix
\section{Difference by Schema Partitioning Strategy \label{sec:diffpartition}}
In this section, we assume that all models used are based on the universal relation with ideal estimation accuracy, and provide a detailed comparison of the methods for obtaining the join cardinality by sampling from the density estimator.
This assumption clarifies what approximations are made outside of the estimation performance of the model for each method.

First, methods based on the universal relation, such as NeuroCard, are always rigorous, as can be seen from \cref{eq:neurocard_cmp}.
On the other hand, DeepDB includes the independent estimation of expected values as shown in \cref{eq:deepdb_cmp}, which indicates that an approximation is made.
Comparing this with the proposed method, we can see that the previous estimation, the $i$th, is limited to $i-1$, but is a conditional expectation, which is a good approximation compared to DeepDB.

\begin{table*}[!tb]\begin{tabular}{c}
$\begin{aligned}
\mathcal{C}_{UR}(Q)
&=
  |J_1|
  \cdot
  \underset{t \sim \mcM_{J_{\bT}}}{\bbE}
  \mqty[
  {\displaystyle \prod_{i=1}^{N}}
    \frac{\textstyle
    \mathbb{I}(\bA, Q_{J_i}, t)
    \cdot
    \prod_{\substack{(T_1, T_2, (A_1, A_2)) \in E_Q \\
    \land T_1 \in V_{J_i} \land T_2 \notin V_{J_i} \land T_2 \in V_{J_{>i}}}}{
        t[F_{(T_1, T_2, (A_1, A_2))}]}
    }{\textstyle\raisebox{-3pt}{$
    \prod_{join \notin E_Q \lor join \notin E_{J_i}}{t[F_{join}]}
    $}}
  ]
  \eqlabel{eq:neurocard_cmp}
\end{aligned}$
\\
$\begin{aligned}
\mathcal{C}_{CBP}(Q)
&=
  |J_1|
  \cdot
  \prod_{i=1}^{N}{
    \underset{t_i \sim \mcM_{J_{\bT}}}{\bbE}
    \mqty[
      \frac{\textstyle
        \mathbb{I}(\bA, Q_{J_i}, t_i)
        \cdot
        \prod_{\substack{(T_1, T_2, (A_1, A_2)) \in E_Q \\
    \land T_1 \in V_{J_i} \land T_2 \notin V_{J_i} \land T_2 \in V_{J_{>i}}}}{
            t_i[F_{(T_1, T_2, (A_1, A_2))}]}
      }{\textstyle\raisebox{-3pt}{$
        \prod_{join \notin E_Q \lor join \notin E_{J_i}}{t_i[F_{join}]}
      $}}
    ]
  }
  \eqlabel{eq:deepdb_cmp}
\end{aligned}$
\\
$\begin{aligned}
\mathcal{C}_{INP}(Q)
&=
  |J_1|
  \cdot
    \prod_{i=1}^{N}
        \underset{t_i \sim \mcM_{J_{\bT}}}{\bbE}
        \Biggl\lbrack
        \frac{\textstyle
        \mathbb{I}(\bA, Q_{J_i}, t)
        \cdot 
        \prod_{
            \substack{(T_1, T_2, (A_1, A_2)) \in E_Q \\
    \land T_1 \in V_{J_i} \land T_2 \notin V_{J_i} \land T_2 \in V_{J_{>i}}}
        }{
            t_i[F_{(T_1, T_2, (A_1, A_2))}]
        }
        }{\textstyle
        \prod_{join \notin E_Q \lor join \notin E_{J_i}}{t_i[F_{join}]}
        }
        \Biggm|
        \Big\langle \FA{j} \in [1,i-1] \land \FA{A} \in (J_i.\bA \cap J_j.\bA) \Big\rangle \land J_i.A = t_j[A]
        \Biggr\rbrack
\eqlabel{eq:ours_cmp}
\end{aligned}$
\end{tabular}\end{table*}

\section{Partitioning Global Schema Graph with Parallel Edges \label{sec:multiedges}}
\begin{figure}[H]
    \centering
    \begin{minipage}[t]{0.49\columnwidth}
    \centering
    \includegraphics[width=\hsize]{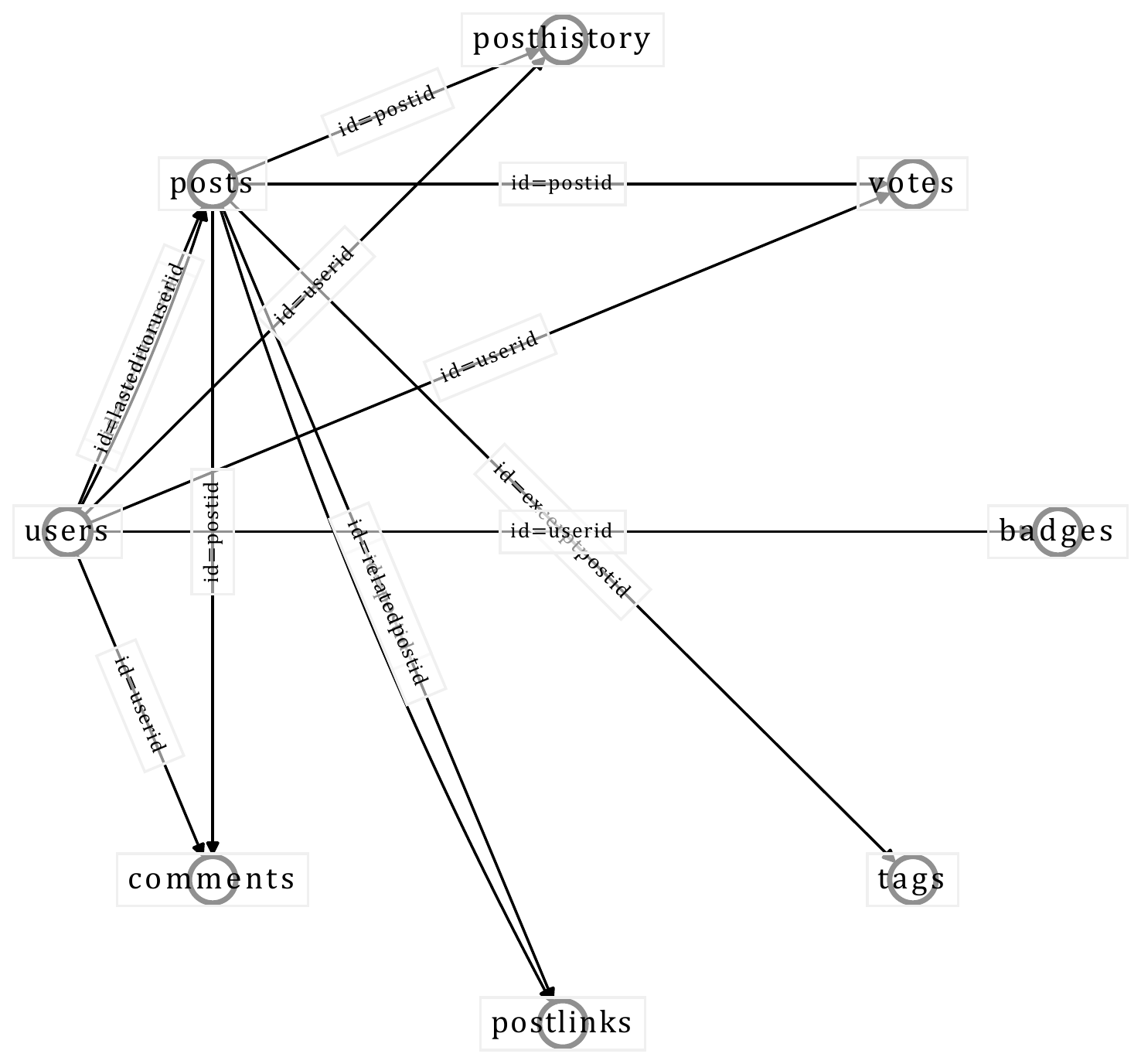}
    \caption{Example of schema graph with multiple edges. \label{fig:g_prop_multi_g}}
    \end{minipage}
    \hspace{0.0\columnwidth}
    \begin{minipage}[t]{0.49\columnwidth}
    \centering
    \includegraphics[width=\hsize]{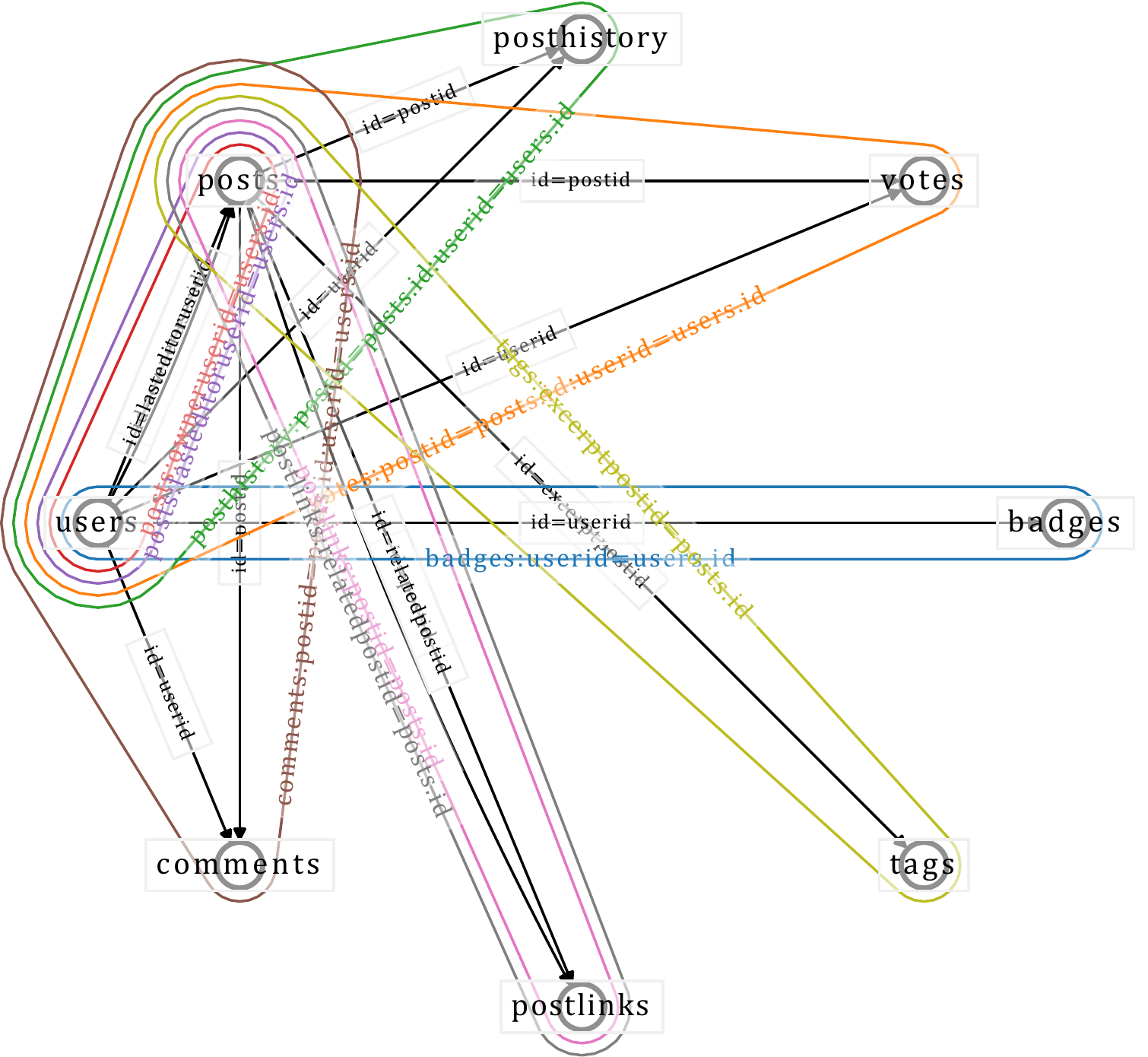}
    \caption{Example of constructing closed in-neighborhood subschemas on schema graph with multiple edges. \label{fig:g_prop_multi_s}}
    \end{minipage}
\end{figure}

In this section, we generalize the schema partitioning strategy described in \cref{ssec:prop_partition} to the case where a global schema graph $G$ is a directed multigraph.
The case with multiple edges is handled edge by edge, since the scaling result of joins differs depending on the edge, i.e., foreign key constraints.
In concrete speaking, closed in-neighborhood subschemas are constructed for each combination of multiple edges.
The procedure is shown in \cref{alg:partitioning}.
First, incoming edges to $v$ are collected for each vertex in the neighborhood (Line~\ref{l:for_ngh_s}~-~\ref{l:for_ngh_e}).
Based on the $edges_list$, which is a collection of edge sets, the direct product sets are obtained (Line~\ref{l:prod}).
Finally, subschemas are constructed for each element of the direct product set (Line~\ref{l:construct_s}).
An example of an actual application for a global schema with multiple edges is shown in \cref{fig:g_prop_multi_g}.
If we focus on \textsf{postlinks} table, there are multiple edges from \textsf{posts} table.
Closed in-neighborhood subschemas are constructed for each combination of multiple edges, resulting in two schemas with different combinations of foreign key constraints between \textsf{postlinks} and \textsf{posts} tables (\cref{fig:g_prop_multi_s}).
If there are multiple edges for multiple neighborhoods, as many subschemas are constructed as the number of combinations.
During inference, subschemas are selected according to join constraints in a query graph.

\begin{algorithm}[H]
\caption{Schema partitioning by closed in-neighborhood considering multiple edges. When multiple edges are included, each subschema is constructed for each pair of edges that is the source of the direct product set. \label{alg:partitioning}}
\small
\begin{algorithmic}[1]
\INPUT{Global schema graph $G_{GLB}:=(V_{GLB},E_{GLB})$}
\OUTPUT{Set of closed in-neighborhood schema graphs $\bm{G}_{CIN}$}
\State{$\bm{G}_{CIN} \gets \{\}$}
\ForEach{$v \in V_{GLB}$}
  \State{$edges\_list \gets \{\}$}
  \ForEach{$u \in N^{-}_{G_{GLB}}(v)$} \label{l:for_ngh_s}
    \State{$edges\_list.append(\{(x,y,c) \in E_{GLB} \mid x=v \land y=u\})$}
  \EndFor \label{l:for_ngh_e}
  \State{$N \gets |edges\_list|$}
  \State{$V_{CIN_v} \gets N^{-}_{G_{GLB}}[v]$}
  \ForEach{$E_{CIN_v} \in \{(e_1,\dots,e_{N}) \mid e_1 \in edges\_list[1] \land \dots \land e_{N} \in edges\_list[N] \}$} \label{l:prod}
    \State{} \Comment{Cartesian product}
    \State{$G_{CIN_v} := (V_{CIN_v},E_{CIN_v})$} \label{l:construct_s}
    \State{$\bm{G}_{CIN}.append(G_{CIN_v})$}
  \EndFor
\EndFor
\State {$\textbf{return} \ \bm{G}_{CIN}$}
\end{algorithmic}
\normalsize
\end{algorithm}

\end{document}